\documentstyle[12pt]{article}

\renewcommand{\theequation}{\arabic{section}.\arabic{equation}}
\newcommand{\cleqn}{\setcounter{equation}{0}}
\newcommand{\sectionnew}[1]{\section{#1}\cleqn}

\newcommand {\Label}[1]{\label{#1}}
\newcommand {\Bibitem}[1]{\bibitem{#1}}

\newcommand{\beq}{\begin{equation}}
\newcommand{\eeq}{\end{equation}}
\newcommand{\beqa}{\begin{eqnarray}}
\newcommand{\eeqa}{\end{eqnarray}}
\newcommand{\beqan}{\begin{eqnarray*}}
\newcommand{\eeqan}{\end{eqnarray*}}
\newcommand{\eqn}[1]{(\ref{#1})}
\newcommand{\eq}[1]{eq.\,(\ref{#1})}
\newcommand{\eqs}[2]{eqs.\,(\ref{#1},\ref{#2})}
\newcommand{\sfrac}[2]{{\textstyle \frac{#1}{#2}}}
\newcommand{\n}{\nonumber \\}
\newcommand{\Winf}{W_{1+\infty}}
\newcommand{\Win}{W_{\infty}}
\newcommand{\Winfp}{W_{1+\infty}[p(D)]}
\newcommand{\WinfMN}{W_{1+\infty}^{M,N}}
\newcommand{\glinf}{\widehat{\mbox{gl}}(\infty)}
\newcommand{\uone}{\widehat{\mbox{u}}(1)}
\newcommand{\Uone}{\mbox{U}(1)}
\newcommand{\tW}{\widetilde{W}}
\newcommand{\bb}{{\bf b}}
\newcommand{\bc}{{\bf c}}
\newcommand{\normal}{{}^{\circ}_{\circ}}
\newcommand{\f}{\flat}
\newcommand{\fb}{\bar{\flat}}
\newcommand{\dket}[1]{\|#1\rangle\!\rangle}
\newcommand{\dbra}[1]{\langle\!\langle#1\|}
\newcommand{\Jc}{{\cal J}}
\newcommand{\Sn}{{\cal S}_n}

\font\csc=cmcsc10 scaled\magstep1
\font\tennn=msbm10
\font\twelvenn=msbm10 scaled\magstep1

\newcommand{\Sbm}[1]{\leavevmode\raise-.15ex\hbox{\twelvenn #1}}
\newcommand{\sbm}[1]{\leavevmode\raise-.15ex\hbox{\tennn #1}}
\newcommand{\bC}{\Sbm{C}}
\newcommand{\bR}{\Sbm{R}}
\newcommand{\bZ}{\Sbm{Z}}
\newcommand{\bz}{\sbm{Z}}

\begin{document}

{\baselineskip=14pt
 \rightline{
 \vbox{\hbox{RIMS-990}
       \hbox{YITP/K-1087}
       \hbox{YITP/U-94-25}
       \hbox{SULDP-1994-7}
       \hbox{August 1994}
}}}

\renewcommand{\thefootnote}{\fnsymbol{footnote}}
\vskip 5mm
\begin{center}
{\large\bf
Representation Theory of The $\Winf$ Algebra\footnote{
Invited talk at ``Quantum Field Theory, Integrable Models and
Beyond", YITP, 14-17 February 1994.
To appear in Progress of Theoretical Physics Proceedings Supplement.
}}

\vspace{10mm}

{\csc Hidetoshi AWATA}\footnote{JSPS fellow}\setcounter{footnote}
{0}\renewcommand{\thefootnote}{\arabic{footnote}}\footnote{
      e-mail address : awata@kurims.kyoto-u.ac.jp},
{\csc Masafumi FUKUMA}\footnote{
      e-mail address : fukuma@yukawa.kyoto-u.ac.jp} ,
{\csc Yutaka   MATSUO}\footnote{
      e-mail address : yutaka@yukawa.kyoto-u.ac.jp}\\
\vskip.1in
and \
{\csc Satoru    ODAKE}\footnote{
      e-mail address : odake@yukawa.kyoto-u.ac.jp}

{\baselineskip=15pt
\it\vskip.25in
  $^1$Research Institute for Mathematical Sciences \\
  Kyoto University, Kyoto 606, Japan \\
\vskip.1in
  $^2$Yukawa Institute for Theoretical Physics \\
  Kyoto University, Kyoto 606, Japan \\
\vskip.1in
  $^3$Uji Research Center, Yukawa Institute for Theoretical Physics \\
  Kyoto University, Uji 611, Japan \\
\vskip.1in
  $^4$Department of Physics, Faculty of Liberal Arts \\
  Shinshu University, Matsumoto 390, Japan
}

\end{center}

\vspace{4mm}

\begin{abstract}
{\baselineskip 15pt
We review the recent development in the
representation theory of the $\Winf$ algebra.
The topics that we concern are,
\begin{itemize}
  \item Quasifinite representation
  \item Free field realizations
  \item (Super) Matrix Generalization
  \item Structure of subalgebras such as $W_\infty$ algebra
  \item Determinant formula
  \item Character formula.
\end{itemize}
}
\end{abstract}

hep-th/9408158
\setcounter{footnote}{0}
\renewcommand{\thefootnote}{\arabic{footnote}}
\newpage
\sectionnew{Introduction}

Symmetry is one of the most important concepts in modern physics,
{\it e.g.} $\mbox{SU}(3)$ symmetry in quark model, gauge symmetry
in gauge theory, conformal symmetry in conformal field theory.
To study physical system from symmetry point of view,
we need the representation theory of the corresponding symmetry
algebra; finite dimensional Lie algebra for quark model or gauge
theory, infinite dimensional Lie algebra (the Virasoro algebra)
for two-dimensional conformal field theory.
Conformal symmetry restricts theories very severely due to its
infinite dimensionality\cite{rBPZ}. In fact, by combining the
knowledge of the representation theory of the Virasoro algebra and
the requirement of the modular invariance, the field contents of
the minimal models were completely classified\cite{rCIZK}.
Another example of the powerfulness of the symmetry argument is
that correlation functions of the XXZ model were determined by
using the representation theory of affine quantum algebra
$U_q\widehat{sl}_2$\cite{rDFJMN}.

When conformal field theory has some extra symmetry, the Virasoro
algebra must be extended,
{\it i.e.} semi-direct products of the Virasoro
algebra with Kac-Moody algebras, superconformal algebras,
the $W$ algebras and parafermions.
The $W_N$ algebra is generated by currents
of conformal spin $2,3,\cdots,N$, and their commutation relation has
non-linear terms\cite{rZFL,rBS}.
The $W$ infinity algebras are Lie algebras obtained
by taking $N\rightarrow\infty$ limit of the $W_N$ algebra.

The $W$ infinity algebras naturally arise in various physical systems,
such as two-dimensional quantum gravity\cite{rFKN,rDVV,rIM,rKS,rSc,rG},
the quantum Hall effects\cite{rCTZ,rIKS},
the membrane\cite{rBST,rFK}, the large $N$ QCD\cite{rGT,rDLS},
and also in the construction of gravitational
instantons\cite{rT,rYC,rP}(see also \cite{rBK2}).
To study these systems we first need to prepare the representation
theory of $W$ infinity algebras, especially the most fundamental
one, the $\Winf$ algebra.

To begin with, we present a short review of the history of the $W$
infinity algebras before the appearance of ref.\cite{rKR}.
By taking an appropriate $N\rightarrow\infty$ limit of the $W_N$
algebra, we can obtain a Lie algebra with infinite number of currents.
Depending on how the background charge scales with $N$, there are
many kinds of $W$ infinity algebras.
The first example is the $w_{\infty}$ algebra\cite{rB}.
Its generators $w^k_n$ ($k,n\in\bZ,k\geq 2$) have the commutation
relation,
\beq
  \Bigl[w^k_n,w^{\ell}_m\Bigr]
  =
  \Bigl((\ell-1)n-(k-1)m\Bigr)w^{k+\ell-2}_{n+m}.
\eeq
$w^2_n$ generates the Virasoro algebra without center and $w^k_n$ has
conformal spin $k$.
This $w_{\infty}$ algebra has a geometrical interpretation as the
algebra of area-preserving diffeomorphisms of two-dimensional phase
space.
However, $w_{\infty}$ admits a central extension only in the
Virasoro sector,
\beq
  \Bigl[w^k_n,w^{\ell}_m\Bigr]
  =
  \Bigl((\ell-1)n-(k-1)m\Bigr)w^{k+\ell-2}_{n+m}
  +\frac{c}{12}(n^3-n)\delta_{n+m,0}\delta^{k\ell}\delta^{k2}.
\eeq
To introduce a central extension in all spin sectors, we must take
another type of the limit $N\rightarrow\infty$
or the deformation of  the $w_{\infty}$ algebra.
By deforming $w_{\infty}$, Pope, Romans and Shen constructed such
algebra, the $\Win$ algebra, in algebraic way by requiring linearity,
closure and the Jacobi identity\cite{rPRS1}.
The $\Win$ algebra is generated by $\tW^k_n$ ($k,n\in\bZ,k\geq 2$) and
its commutation relation is given
by
\beqa
  \Bigl[\tW^k_n,\tW^{\ell}_m\Bigr]
  &\!\!=\!\!&
  \sum_{r=0}^{\infty}\tilde{g}^{k\ell}_{2r}(n,m)\tW^{k+\ell-2-2r}_{n+m}
  \n
  &&
  +\tilde{c}\delta^{k\ell}\delta_{n+m,0}
  \frac{1}{k-1}{2(k-1)\choose k-1}^{-1}{2k\choose k}^{-1}
  \prod_{j=-(k-1)}^{k-1}(n+j),
  \Label{PRSWin}
\eeqa
where $\tilde{c}$ is the central charge of the Virasoro algebra
generated by $\tW^2_n$, and the structure constant
$\tilde{g}^{k\ell}_r$ is given by
\beqa
  \tilde{g}^{k\ell}_r(n,m)
  &\!\!=\!\!&
  \frac{1}{2^{2r+1}(r+1)!}\phi^{k\ell}_r(0,0)N^{k,\ell}_r(n,m), \\
  N^{x,y}_r(n,m)
  &\!\!=\!\!&
  \sum_{s=0}^{r+1}(-1)^s {r+1\choose s}
  \lbrack x-1+n \rbrack_{r+1-s}
  \lbrack x-1-n \rbrack_s \n
  && \hspace{27mm}
  \times
  \lbrack y-1-m \rbrack_{r+1-s}
  \lbrack y-1+m \rbrack_s, \\
  \phi^{k\ell}_r(x,y)
  &\!\!=\!\!&
  {}_4F_{\,3} \Biggl[
  \begin{array}{c}
    -\frac{1}{2}-x-2y,\frac{3}{2}-x+2y,
    -\frac{r+1}{2}+x,-\frac{r}{2}+x \\
    -k+\frac{3}{2},-\ell+\frac{3}{2},k+\ell-r-\frac{3}{2}
  \end{array}
  ;1 \Biggr], \\
  {}_4F_{\,3} \biggl[
  \begin{array}{c}
    a_1,a_2,a_3,a_4 \\
    b_1,b_2,b_3
  \end{array}
  ;z \biggr]
  &\!\!=\!\!&
  \sum_{n=0}^{\infty}
  \frac{(a_1)_n(a_2)_n(a_3)_n(a_4)_n}
       {(b_1)_n(b_2)_n(b_3)_n}
  \frac{z^n}{n!}, \\
  \lbrack x\rbrack_n
  =\prod_{j=0}^{n-1}(x-j),
  &&
  (x)_n=\prod_{j=0}^{n-1}(x+j),\quad
  {x \choose n}=\frac{[x]_n}{n!}.
\eeqa
The $w_{\infty}$ algebra is obtained from $\Win$ by contraction;
we take the $q\rightarrow 0$ limit after rescaling
$\tW^k_n\rightarrow q^{2-k}\tW^k_n$.
Furthermore they constructed the $\Winf$ algebra which contains a spin 1
current\cite{rPRS2}. The $\Winf$ algebra is generated by $W^k_n$
($k,n\in\bZ,k\geq 1$) and its commutation relation is given by
\beqa
  \Bigl[W^k_n,W^{\ell}_m\Bigr]
  &\!\!=\!\!&
  \sum_{r=0}^{\infty}g^{k\ell}_{2r}(n,m)W^{k+\ell-2-2r}_{n+m} \n
  &&
  +c\delta^{k\ell}\delta_{n+m,0}
  \frac{2}{k}{2(k-1)\choose k-1}^{-1}{2k\choose k}^{-1}
  \prod_{j=-(k-1)}^{k-1}(n+j),
  \Label{PRSW}
\eeqa
where $c$ is the central charge of the Virasoro algebra
generated by $W^2_n$, and the structure constant
$g^{k\ell}_r$ is given by
\beq
  g^{k\ell}_r(n,m)
  =
  \frac{1}{2^{2r+1}(r+1)!}\phi^{k\ell}_r(0,-\sfrac{1}{2})
  N^{k,\ell}_r(n,m).
\eeq
Since $\tilde{g}^{k\ell}_r(n,m)=0$ for $k+\ell-r<4$
and $g^{k\ell}_r(n,m)=0$ for $k+\ell-r<3$,
the summations over $r$ are finite sum and the algebras close.
These commutation relations are consistent with the hermitian
conjugation $\tW^{k\dagger}_n=\tW^k_{-n}$,
$W^{k\dagger}_n=W^k_{-n}$, and have diagonalized central terms.
The $\Winf$ algebra contains the $\Win$ algebra as a
subalgebra\cite{rPRS3}, but it is nontrivial in these basis.
Moreover various extensions were constructed;
super extension ($W^{1,1}_{\infty}$)\cite{rBPRSS,rBdWV}, $\mbox{u}(M)$
matrix version of $\Win$ ($W^M_{\infty}$)\cite{rBK},
$\mbox{u}(N)$ matrix version of $\Winf$ ($W^N_{1+\infty}$)\cite{rOS},
and they were unified as $W^{M,N}_{\infty}$\cite{rO}.
Based on the coset model $\mbox{SL}(2,\bR)_k/\Uone$, a nonlinear
deformation of $\Win$, $\widehat{W}_{\infty}(k)$, was also
constructed\cite{rBK3}.

When we study the representation theory of $W$ infinity algebras,
we encounter the difficulty that infinitely many states possibly
appear at each energy level, reflecting the infinite number of
currents. For example, even at level 1, there are infinite number
of states $W^k_{-1}|\mbox{hws}\rangle$ ($k=1,2,3,\cdots$) for
generic representation, so we could not treat these states,
{\it e.g.} computation of the Kac determinant. Moreover they are
not the simultaneous eigenstates of the Cartan generators $W^k_0$
($k=1,2,\cdots$).
Only restricted class of the representation were studied by using
$\bZ_{\infty}$ parafermion and coset model\cite{rBK} or
free field realizations\cite{rO}. In the free field realization,
there are only finite number of states at each energy level
because the number of oscillators is finite at each level.

Last year Kac and Radul overcame this difficulty of
infiniteness\cite{rKR}. They proposed the quasifinite representation,
which has only finite number of states at each energy level, and
studied this class of representations in detail.
{}From physicist point of view, this notion is the abstraction of
the property that the free field realizations have.

In this article, we would like to review the recently developed
representation theory of the $W$ infinity algebras, mainly the $\Winf$
algebra\cite{rKR,rM,rAFOQ,rAFMO1,rAFMO2,rAFMO3,rAFMO4,rFKRW}.
In section 2 we give the definition of the $\Winf$ algebra and
its (super)matrix generalizations.
Various subalgebras of $\Winf$ are also given.
In section 3 free field realizations of $\Winf$ and $\WinfMN$ are given.
Using these we derive the full character formulae for those
representations.
In section 4 the quasifinite representation is introduced, and its
general properties are presented.
In section 5, after describing the Verma module, we compute the
Kac determinant at lower levels for some representations
(its results are given in appendix A).
On the basis of this computation we derive the analytic form of the Kac
determinant and the full character formulae.
Appendix B is devoted to the description of the Schur function.

\sectionnew{$W$ infinity algebras}

In this section we define the $\Winf$ algebra and its (super)matrix
generalization $\WinfMN$.
We also give a systematic method to construct a family of subalgebras
of $\Winf$.

\subsection{The $\Winf$ algebra}

Since the $W$ algebras were originally introduced as
extensions of the Virasoro algebra,
we first recall the Virasoro algebra.
Let us consider the Lie algebra of the diffeomorphism group
on the circle whose coordinate is $z$. The generator of this
Lie algebra is $l_n=-z^{n+1}\frac{d}{dz}$ and its commutation
relation is
$$
  [l_n,l_m]=(n-m)l_{n+m}.
$$
The Virasoro algebra, whose generators are denoted as $L_n$,
is the central extension of this algebra,
$$
  [L_n,L_m]=(n-m)L_{n+m}+\frac{c}{12}(n^3-n)\delta_{n+m,0}.
$$

Besides $l_n$, we may consider the higher order differential operators
on the circle, $z^n\bigl(\frac{d}{dz}\bigr)^m$ ($n,m\in\bZ,m\geq 0$).
Instead of $z^n\bigl(\frac{d}{dz}\bigr)^m$, we take a basis
$z^nD^k$ ($n,k\in\bZ,k\geq 0$) with $D=z\frac{d}{dz}$.
Since $f(D)z^n=z^nf(D+n)$, the commutation relation of the
differential operators is
\beq
  \Bigl[z^nf(D),z^mg(D)\Bigr]
  =
  z^{n+m}f(D+m)g(D)-z^{n+m}f(D)g(D+n),
\eeq
where $f$ and $g$ are polynomials. The $\Winf$ algebra is the central
extension of this Lie algebra of differential operators on the
circle\cite{rPRS3,rBKK,rKR}.
We denote the corresponding generators by $W(z^nD^k)$ and the central
charge by $C$. The commutation relation is\cite{rKR}
\beqa
  &&\Bigl[ W(z^nf(D)), W(z^mg(D))\Bigr] \n
  &\!\!=\!\!&
  W(z^{n+m}f(D+m)g(D))
  -W(z^{n+m}f(D)g(D+n))
  +C \Psi(z^nf(D),z^mg(D)).
\eeqa
Here the 2-cocycle $\Psi$ is defined by
\beqa
  &&\Psi(z^nf(D),z^mg(D)) \n
  &\!\!=\!\!&\delta_{n+m,0}
  \biggl( \theta(n\geq 1) \sum_{j=1}^n f(-j)g(n-j)
        -\theta(m\geq 1) \sum_{j=1}^m f(m-j)g(-j)
  \biggr),
\eeqa
where $\theta(P)=1$ (or 0) when the proposition $P$ is true (or false).
The 2-cocycle is unique up to coboundary\cite{rKPWBMSFGL}.
By introducing $z^n e^{xD}$ as a generating series for $z^n D^k$,
the above 2-cocycle and commutation relation can be rewritten in a
simpler form:
\beqa
  \Psi(z^ne^{xD},z^me^{yD})
  &\!\!=\!\!&
  -\frac{e^{mx}-e^{ny}}{e^{x+y}-1}\delta_{n+m,0},\\
  \Bigl[ W(z^ne^{xD}), W(z^me^{yD})\Bigr]
  &\!\!=\!\!&
  \left(e^{mx}-e^{ny} \right)
   W(z^{n+m}e^{(x+y)D})
  -C \frac{e^{mx}-e^{ny}}{e^{x+y}-1}\delta_{n+m,0}.
  \Label{Winf}
\eeqa

Since $\Winf$ is a Lie algebra, we can take any invertible linear
combination of $W(z^nD^k)$ as a basis.
The basis $W^k_n$ in section 1, \eq{PRSW}, is expressed as
\beqa
  W^{k+1}_n
  &\!\!=\!\!&
  W(z^n f^k_n(D))\quad (k\geq 0),\quad c=C, \n
  f^k_n(D)
  &\!\!=\!\!&
  {2k \choose k}^{-1}
  \sum_{j=0}^k(-1)^j{k \choose j}^2
  \lbrack -D-n-1\rbrack_{k-j} \lbrack D\rbrack_j
  =(-1)^kD^k+\cdots.
  \Label{fkn}
\eeqa

$\Winf$ contains the $\uone$ subalgebra generated by
$J_n=W(z^n)$ and the Virasoro subalgebra generated by
$L_n=-W(z^nD)$ with the central charge $c_{Vir}=-2C$.
$L_0$ counts the energy level; $[L_0,W(z^nf(D))]=-nW(z^nf(D))$.
We will regard $W(z^nf(D))$ with $n>0$ ($n<0$) as annihilation
(creation) operators, respectively.
The Cartan subalgebra of $\Winf$ is generated by $W(D^k)$ ($k\geq 0$),
so it is infinite dimensional.
$W^2_n=L_n-\frac{n+1}{2}J_n$ also generates the Virasoro algebra
with $c_{Vir}=C$. Moreover there are two one-parameter families of the
Virasoro subalgebras generated by\cite{rFKRW}
\beq
  L_n-(\alpha n+\beta)J_n,\quad
  (\alpha=\beta,1-\beta;\beta\in\bC),
\eeq
whose central charge is
\beq
  c_{Vir}=2(-1+6\beta-6\beta^2)C.
\eeq
The $\uone$ current $J_n$ is anomalous except
for $\beta=\frac{1}{2}$.

Since $\Winf$ contains the $\uone$ subalgebra, $\Winf$ has a
one-parameter family of automorphisms which we call the spectral
flow\cite{rSS,rBPRSS}. The transformation rule is given by\cite{rAFMO1}
\beq
  W'(z^ne^{xD})=
  W(z^ne^{x(D+\lambda)})
  -C\frac{e^{\lambda x}-1}{e^x-1}\delta_{n0},
  \Label{sf}
\eeq
where $\lambda\in\bC$ is an arbitrary parameter.
For lower components, for example, it is expressed as
\beqa
  J'_n&\!\!=\!\!&J_n-\lambda C\delta_{n0},\n
  L'_n&\!\!=\!\!&L_n-\lambda J_n
  +\sfrac{1}{2}\lambda(\lambda-1)C\delta_{n0}.
\eeqa
One can easily check that
new generator $W'(\cdot)$ satisfies the same commutation relation
as the original one $W(\cdot)$, \eq{Winf}.

The Hermitian conjugation $\dagger$ is defined by
\beq
  W(z^nD^k)^{\dagger}=W(z^{-n}(D-n)^k),
  \Label{dagger}
\eeq
and $(aA+bB)^{\dagger}=\bar{a}A^{\dagger}+\bar{b}B^{\dagger}$,
$(AB)^{\dagger}=B^{\dagger}A^{\dagger}$.
The commutation relation \eq{Winf} is invariant under $\dagger$.
$f^k_n(D)$, \eq{fkn}, satisfies $f^k_n(D-n)=f^k_{-n}(D)$,
which implies $W^{k\dagger}_n=W^k_{-n}$.

Finally we remark that $\Winf$ is generated by $W(z^{\pm 1})$ and
$W(D^2)$, namely $W(z^nD^k)$ is expressed as a commutator of
$W(z^{\pm 1})$ and $W(D^2)$.

\subsection{(Super)Matrix generalization of $\Winf$}

We can construct a (super)matrix generalization of $\Winf$.
Let us consider the $(M+N)\times(M+N)$ supermatrices
$\mbox{M}(M|N;\bC)$. An element of $\mbox{M}(M|N;\bC)$ has the
following form:
\beq
  A=
  \left(
  \begin{array}{cc}
    A^{(0)}&A^{(+)} \\
    A^{(-)}&A^{(1)} \\
  \end{array}
  \right),
\eeq
where $A^{(0)},A^{(1)},A^{(+)},A^{(-)}$ are $M\times M$, $N\times N$,
$M\times N$, $N\times M$ matrices, respectively, with complex entries.
$\bZ_2$-gradation is denoted by $|A|$; $|A|=0$ for $\bZ_2$-even and
$|A|=1$ for $\bZ_2$-odd. $A^{(0)}$ and $A^{(1)}$ are $\bZ_2$-even and
$A^{(+)}$ and $A^{(-)}$ are $\bZ_2$-odd. $\bZ_2$-graded commutator
is
\beq
  [A,B\}=AB-(-1)^{|A||B|}BA.
\eeq
The supertrace is
\beq
  \mbox{str}\,A
  =
  \mbox{tr}\,A^{(0)}-\mbox{tr}\,A^{(1)},
\eeq
and satisfies $\mbox{str}\,(AB)=(-1)^{|A||B|}\mbox{str}\,(BA)$.

$\mbox{M}(M|N;\bC)$ generalization of $\Winf$,
whose generators are $W(z^nD^kA)$
($n,k\in\bZ,k\geq 0,A\in\mbox{M}(M|N;\bC)$) and the center $C$,
is defined by the following (anti-)commutation relation:
\beqa
  &&\Bigl[ W(z^nf(D)A), W(z^mg(D)B)\Bigr\} \n
  &\!\!=\!\!&
  W(z^{n+m}f(D+m)g(D)AB)
  -(-1)^{|A||B|}W(z^{n+m}f(D)g(D+n)BA) \n
  && -C \Psi(z^nf(D),z^mg(D)) {\rm str}(AB).
\eeqa
We call this ($\bZ_2$-graded) Lie algebra the $\WinfMN$ algebra,
which satisfies the Jacobi identity
\beqa
  &&
  (-1)^{|A_1||A_3|}
  \Bigl[ W(z^{n_1}f_1(D)A_1),
  \Bigl[ W(z^{n_2}f_2(D)A_2),
  W(z^{n_3}f_3(D)A_3)\Bigr\}\Bigr\} \n
  && \hspace{80mm} + \mbox{ cyclic permutation} =0.
\eeqa
The original $\Winf$ algebra corresponds to $M=0$, $N=1$.
$M=0$ case was constructed in \cite{rOS}, and $M=N=1$ case
in \cite{rAFMO2}.

The $\WinfMN$ algebra contains $\mbox{M}(M|N;\bC)$ current algebra
generated by $W(z^nA)$. For $M=0$, it is the $\widehat{\mbox{gl}}(N)$
(or $\widehat{\mbox{u}}(N)$) algebra with level $C$.
Since $\WinfMN$ contains $M+N$ $\uone$ subalgebras,
$\WinfMN$ has $(M+N)$-parameter family of automorphisms (spectral
flow). Its transformation rule is
\beqa
  W'(z^ne^{xD}E^{(0)}_{ab})
  &\!\!=\!\!&
  W(z^{n-\mu^a+\mu^b}e^{x(D+\mu^b)}E^{(0)}_{ab})
  +C\frac{e^{\mu^ax}-1}{e^x-1}\delta_{ab}\delta_{n0}, \n
  W'(z^ne^{xD}E^{(1)}_{ij})
  &\!\!=\!\!&
  W(z^{n-\lambda^i+\lambda^j}e^{x(D+\lambda^j)}E^{(1)}_{ij})
  -C\frac{e^{\lambda^ix}-1}{e^x-1}\delta_{ij}\delta_{n0}, \n
  W'(z^ne^{xD}E^{(+)}_{aj})
  &\!\!=\!\!&
  W(z^{n-\mu^a+\lambda^j}e^{x(D+\lambda^j)}E^{(+)}_{aj}), \n
  W'(z^ne^{xD}E^{(-)}_{ib})
  &\!\!=\!\!&
  W(z^{n-\lambda^i+\mu^b}e^{x(D+\mu^b)}E^{(-)}_{ib}),
\eeqa
where $\mu^a$ ($a=1,\cdots,M$) and $\lambda^i$ ($i=1,\cdots,N$) are
arbitrary parameters, and $E^{(\alpha)}_{pq}$ is a matrix unit,
$(E^{(\alpha)}_{pq})_{p'q'}=\delta_{pp'}\delta_{qq'}$.

$L_n=-W(z^nD\cdot 1)$ generates the Virasoro algebra with the central
charge $c_{Vir}=2(M-N)C$. $L_0$ counts the energy level.
The Cartan subalgebra of $\WinfMN$ is generated by
$W(D^kE^{(0)}_{aa})$ ($k\geq 0,a=1,\cdots,M$) and
$W(D^kE^{(1)}_{ii})$ ($k\geq 0,i=1,\cdots,N$).

\subsection{Subalgebras of $\Winf$}

Although $\Winf$ was constructed from $\Win$ by adding a spin-1
current historically, it is natural to regard that $\Win$ is
obtained from $\Winf$ by truncating a spin-1 current\cite{rPRS3}.
The higher spin truncation of $\Winf$ was also constructed\cite{rBKK}.
We will give a systematic method to construct a family of subalgebras
of the $\Winf$ algebra\cite{rAFMO4}.

Let us choose a polynomial $p(D)$ and set
\beq
  \tW(z^nD^k)=W(z^nD^kp(D)), \quad (n,k\in\bZ,k\geq 0).
\eeq
Then commutator of $\tW(z^nD^k)$ closes:
\beqa
  &&\Bigl[\tW(z^nf(D)),\tW(z^mg(D))\Bigr] \n
  &\!\!=\!\!&
  \tW(z^{n+m}f(D+m)g(D)p(D+m))
  -\tW(z^{n+m}f(D)g(D+n)p(D+n)) \n
  &&+C\Psi(z^nf(D)p(D),z^mg(D)p(D)),
\eeqa
or equivalently
\beqa
  \left[ \tW(z^n e^{xD}), \tW(z^m e^{yD})\right]
  &\!\!=\!\!&
  \left( p(\sfrac{d}{dx}) e^{mx}
         -p(\sfrac{d}{dy}) e^{ny} \right)
   \tW(z^{n+m}e^{(x+y)D}) \n
  &&
  -C p(\sfrac{d}{dx}) p(\sfrac{d}{dy})
  \frac{e^{mx}-e^{ny}}{e^{x+y}-1}\delta_{n+m,0}.
\eeqa
We call this subalgebra $\Winfp$.

In this subalgebra there are no currents with spin $\leq\deg p(D)$.
The $\Win$ algebra corresponds to the choice $p(D)=D$.
The basis $\tW^k_n$ \eq{PRSWin} is expressed as
\beqa
  \tW^{k+2}_n
  &\!\!=\!\!&
  \tW (z^n \tilde{f}^k_n(D))\quad (k\geq 0), \n
  \tilde{f}^k_n(D)
  &\!\!=\!\!&
  -{2(k+1) \choose k+1}^{-1}
  \sum_{j=0}^k(-1)^j {k \choose j} {k+2 \choose j+1}
  \lbrack -D-n-1 \rbrack_{k-j} \lbrack D-1 \rbrack_j \n
  &\!\!=\!\!&
  (-1)^{k-1}D^k+\cdots.
\eeqa
We remark that the Virasoro generators exist only if $\deg p(w) \leq 1$.
In the case of $\Win$, the Virasoro generator $L_n$ is given
by $L_n=-\tW (z^n)$ whose central charge, $\tilde{c}_{Vir}$, is
related to $C$ as $\tilde{c}_{Vir}=-2C$ \cite{rPRS3}.
For $\deg p(w) \geq 2$, we extend the algebra introducing the $L_0$
operator such as to count the energy level,
$\left[ L_0, \tW (z^n f(D)) \right]$ $=$ $-n\tW (z^n f(D))$.

Next we give another type of subalgebra of $\Winf$.
For any positive integer $p$, $\Winf$ with the central charge $C$
contains $\Winf$ with the central charge $pC$\cite{rFKN}.
We denote its generator by $\bar{W}(z^nD^k)$ ($n,k\in\bZ,k\geq 0$).
$\bar{W}(\cdot)$ is given by
\beqa
  \bar{W}(z^ne^{xD})
  &\!\!=\!\!&
  W(z^{pn}e^{x\frac{1}{p}D})
  -C\biggl(\frac{1}{e^{\frac{1}{p}x}-1}
          -\frac{p}{e^x-1}\biggr)\delta_{n0} \n
  &\!\!=\!\!&
  W(z^{pn}e^{x\frac{1}{p}D})
  -C\sum_{j=0}^{p-1}\frac{e^{\frac{j}{p}x}-1}{e^x-1}
  \delta_{n0}.
  \Label{pC}
\eeqa
Essentially this is interpreted as the change of variable, $\zeta=z^p$,
$\zeta\frac{d}{d\zeta}=\frac{1}{p}D$.

For $\WinfMN$, these type of subalgebras
such as $W^M_{\infty}$\cite{rBK} and $W^{M,N}_{\infty}$\cite{rO,rAFMO2}
can be treated similarly.

\sectionnew{Free field realizations}

In this section we give the free field realizations of
$\Winf$ and $\WinfMN$. Using these realizations, we give
their full character formulae.

\subsection{$\Winf$}

The $\Winf$ algebra is known to be realized by free fermion\cite{rBPRSS}
or $\bb\bc$ ghost\cite{rM}
\beqa
  &&
  \bb(z)=\sum_{r\in\bz}\bb_rz^{-r-\lambda-1}, \quad
  \bc(z)=\sum_{s\in\bz}\bc_sz^{-s+\lambda}, \quad
  \bb(z)\bc(w)\sim\frac{\epsilon}{z-w}, \n
  &&
  \bb_r|\lambda\rangle=
  \bc_s|\lambda\rangle=0\;(r\geq 0,s\geq 1),\quad
  \bc_s^{\dagger}=\bb_{-s},
\eeqa
where $\epsilon=1$ for fermionic ghost $bc$ or
$\epsilon=-1$ for bosonic ghost $\beta\gamma$.
The $\Winf$ algebra with $C=\epsilon$ is realized by sandwiching
a differential operator between $\bb\bc$:
\beqa
  W(z^ne^{xD})
  &\!\!=\!\!&
  \oint\frac{dz}{2\pi i}
  \normal\bb(z)z^ne^{xD}\bc(z)\normal \n
  &\!\!=\!\!&
  \oint\frac{dz}{2\pi i}
  :\bb(z)z^ne^{xD}\bc(z):
  -\epsilon\frac{e^{\lambda x}-1}{e^x-1}\delta_{n0} \n
  &\!\!=\!\!&
  \sum_{{\scriptstyle r,s}\in\bz \atop {\scriptstyle r+s=n}}
  e^{x(\lambda-s)}E(r,s)
  -\epsilon\frac{e^{\lambda x}-1}{e^x-1}\delta_{n0}.
  \Label{Wbc}
\eeqa
Here the normal ordering $\normal\quad\normal$ means subtracting the
singular part and another normal ordering $:\bb_r\bc_s:$ means
$\bb_r\bc_s$ if $r\leq -1$ and $\epsilon\bc_s\bb_r$ if $r\geq 0$.
$E(r,s)$ is defined by
\beq
  E(r,s)=\;:\bb_r\bc_s:,
\eeq
and generates the $\glinf$ algebra:
\beqa
  \Bigl[E(r,s),E(r',s')\Bigr]
  &\!\!=\!\!&
  \delta_{r'+s,0}E(r,s')-\delta_{r+s',0}E(r',s) \n
  &&
  +C\delta_{r+s',0}\delta_{r'+s,0}
  \Bigl(\theta(r\geq 0)-\theta(r'\geq 0)\Bigr),
  \Label{glinf}
\eeqa
where $C=\epsilon$ in this case.
We remark that the spectral flow transformation \eq{sf} with
parameter $\lambda'$ is obtained by replacing $\bb,\bc$ in \eq{Wbc}
with $\bb'(z)=z^{-\lambda'}\bb(z)$, $\bc'(z)=z^{\lambda'}\bc(z)$.

{}From \eq{Wbc} we obtain
\beqa
  W(z^nD^k)|\lambda\rangle
  &\!\!=\!\!&
  0 \quad (n\geq 1,k\geq 0), \n
  -W(e^{xD})|\lambda\rangle
  &\!\!=\!\!&
  \epsilon\frac{e^{\lambda x}-1}{e^x-1}|\lambda\rangle.
\eeqa
This means that $|\lambda\rangle$ is the highest weight state
of $\Winf$ and its weight is
\beq
  W(D^k)|\lambda\rangle
  =
  \epsilon\Delta^{\lambda}_k|\lambda\rangle,
\eeq
where $\Delta^{\lambda}_k$ is the Bernoulli polynomial defined by
\beq
  -\frac{e^{\lambda x}-1}{e^x-1}
  =
  \sum_{k=0}^{\infty}\Delta^{\lambda}_k\frac{x^k}{k!}.
  \Label{dl}
\eeq

To express how many states exist in the simultaneous eigenspace
of the Cartan generators $W(D^k)$, the full character formula is
introduced as
\beq
  \mbox{ch}=\mbox{tr}\,
  \exp\biggl(\sum_{k=0}^{\infty}g_kW(D^k)\biggr),
  \Label{ch}
\eeq
where the trace is taken over the irreducible representation space
and $g_k$ are parameters.
The states in the representation space are linear combinations of
the following states:
$$
  W(z^{-n_1}D^{k_1})\cdots W(z^{-n_m}D^{k_m})|\lambda\rangle.
$$
This state, however, is not the simultaneous eigenstate of $W(D^k)$,
because
\beq
  \Bigl[W(D^k),W(z^{-n}f(D))\Bigr]
  =
  W(z^{-n}((D-n)^k-D^k)f(D)).
  \Label{WDkW}
\eeq
On the other hand, the states in the Fock space of $\bb\bc$ ghosts
are linear combinations of the following states:
$$
  \bb_{-r_1}\cdots\bb_{-r_k}
  \bc_{-s_1}\cdots\bc_{-s_{\ell}}|\lambda\rangle,
$$
which are simultaneous eigenstates of $W(D^k)$,
because
\beq
  \Bigl[ W(D^k),\bb_{-r}\Bigr]
  =(\lambda-r)^k\bb_{-r},\quad
  \Bigl[ W(D^k),\bc_{-s}\Bigr]
  =-(\lambda+s)^k\bc_{-s}.
\eeq

Using this property, we derive the full character
formula\cite{rAFOQ,rAFMO1}.
For the fermionic case ($\epsilon=1$), it is well known that
the fermion Fock space can be decomposed into the irreducible
representation spaces of $\uone$ current algebra
(cf.~\eq{N}). Since $\Winf$-generator does not change the
$\Uone$--charge and $\Winf$ contains $\uone$ as a subalgebra,
each $\uone$ representation space is also the representation
space of $\Winf$ and irreducible\cite{rO,rAFOQ}.
For the bosonic case ($\epsilon=-1$),
the sector of vanishing $\Uone$--charge in the Fock space
is the irreducible representation space
of $\Winf$ \cite{rAFMO1}.
By taking a trace over whole Fock space,
we define $S^{\lambda;\epsilon}_m$ as follows:
\beq
  \sum_{m\in\bz}S^{\lambda;\epsilon}_mt^{-m}
  =
  e^{\epsilon\sum_{k=0}^{\infty}g_k\Delta^{\lambda}_k}
  \prod_{r=1}^{\infty}
  \Bigl(1+\epsilon tu_r(\lambda)\Bigr)^{\epsilon}
  \prod_{s=0}^{\infty}
  \Bigl(1+\epsilon t^{-1}v_s(\lambda)\Bigr)^{\epsilon},
  \Label{Sgen}
\eeq
where $t$ counts the $\Uone$--charge and $u_r(\lambda)$,
$v_s(\lambda)$ are
\beq
  u_r(\lambda)=e^{\sum_{k=0}^{\infty}g_k(\lambda-r)^k},\quad
  v_s(\lambda)=e^{-\sum_{k=0}^{\infty}g_k(\lambda+s)^k}.
  \Label{uv}
\eeq
Then the full character for $|\lambda\rangle$ is given by
\beq
  \mbox{ch }=S^{\lambda;\epsilon}_0.
\eeq
We remark that $S^{\lambda;1}_m=S^{\lambda+m;1}_0$ by the above
statement. So we abbreviate $S_{\lambda}=S^{\lambda;1}_0$.

Products in \eq{Sgen} can be written as
\beq
  \prod_{r=1}^{\infty}
  \Bigl(1+\epsilon tu_r(\lambda)\Bigr)^{\epsilon}
  =
  e^{-\epsilon\sum_{\ell=1}^{\infty}
     x_{\ell}(\lambda)(-\epsilon t)^{\ell}},\quad
  \prod_{s=0}^{\infty}
  \Bigl(1+\epsilon t^{-1}v_s(\lambda)\Bigr)^{\epsilon}
  =
  e^{-\epsilon\sum_{\ell=1}^{\infty}
     y_{\ell}(\lambda)(-\epsilon t)^{-\ell}},
\eeq
where
\beq
  x_{\ell}(\lambda)=
  \frac{1}{\ell}\sum_{r=1}^{\infty}u_r(\lambda)^{\ell},\quad
  y_{\ell}(\lambda)=
  \frac{1}{\ell}\sum_{s=0}^{\infty}v_s(\lambda)^{\ell}.
  \Label{xy}
\eeq
By introducing the elementary Schur polynomials $P_n$ (see
appendix B, \eq{Pn}),
$S^{\lambda;\epsilon}_m$ is expressed as
\beq
  S^{\lambda;\epsilon}_m
  =
  (-\epsilon)^me^{\epsilon\sum_{k=0}^{\infty}g_k\Delta^{\lambda}_k}
  \sum_{a\in\bz}P_a(-\epsilon x(\lambda))P_{a+m}(-\epsilon y(\lambda)).
  \Label{SPP}
\eeq

To understand the full character formula, we specialize the parameters
$g_k$ as
\beq
  g_k=-2\pi i\tau\delta_{k1},\quad (q=e^{2\pi i\tau}),
  \Label{gk}
\eeq
which correspond to $\mbox{tr}\,q^{L_0}$.
For this choice, \eq{xy} becomes
\beq
  x_{\ell}(\lambda)
  =\frac{1}{\ell}\frac{q^{(1-\lambda)\ell}}{1-q^{\ell}},\quad
  y_{\ell}(\lambda)
  =\frac{1}{\ell}\frac{q^{\lambda\ell}}{1-q^{\ell}}.
\eeq
Then the specialized character is given by
\beqa
  S_{\lambda}
  &\!\!=\!\!&
  q^{\frac{1}{2}\lambda(\lambda-1)}
  \sum_{m=0}^{\infty}q^{m^2}\prod_{j=1}^m\frac{1}{(1-q^j)^2}
  \Label{S1a} \\
  &\!\!=\!\!&
  q^{\frac{1}{2}\lambda(\lambda-1)}
  \prod_{j=1}^{\infty}\frac{1}{1-q^j},
  \Label{S1b} \\
  S^{\lambda;-1}_0
  &\!\!=\!\!&
  q^{-\frac{1}{2}\lambda(\lambda-1)}
  \sum_{m=0}^{\infty}q^m\prod_{j=1}^m\frac{1}{(1-q^j)^2}
  \Label{Sm1a} \\
  &\!\!=\!\!&
  q^{-\frac{1}{2}\lambda(\lambda-1)}
  \prod_{j=1}^{\infty}\frac{1}{(1-q^j)^2}
  \cdot
  \sum_{m=0}^{\infty}(-1)^mq^{\frac{1}{2}m(m+1)}.
  \Label{Sm1b}
\eeqa
Eqs.\,(\ref{S1a},\ref{Sm1a}) are derived by \eq{SPP} and
\eqs{qPnr}{qPnc} in Appendix B.
Eqs.\,(\ref{S1b},\ref{Sm1b}) are derived by \eq{Sgen} and
Jacobi's triple product identity or characters of $\Win$
with $c=2$ \cite{rO,rAFMO1}.

By tensoring the above free field realizations, we obtain free
field realizations of $\Winf$ with integer $C$;
\beq
  C=\sum_i\epsilon_i,\quad
  -W(e^{xD})|\lambda\rangle
  =
  \sum_i\epsilon_i\frac{e^{\lambda_ix}-1}{e^x-1}|\lambda\rangle.
  \Label{tensor}
\eeq
However, the character for this representation can not be obtained
by the method given in this section.
We will give another method in section 5.

Free field realization of $\Winfp$ is obtained from that of $\Winf$
\cite{rAFMO4}.

\subsection{$\WinfMN$}

Results in the previous section are generalized easily\cite{rAFMO2}.
Let us introduce $M$ pairs of $\beta\gamma$ ghosts
and $N$ pairs of $bc$ ghosts:
\beqa
  &&
  \beta^a(z)=\sum_{r\in\bz}\beta^a_rz^{-r-\mu_a-1}, \quad
  \gamma^a(z)=\sum_{s\in\bz}\gamma^a_sz^{-s+\mu_a}, \quad
  (a=1,\cdots,M), \n
  &&
  b^i(z)=\sum_{r\in\bz}b^i_rz^{-r-\lambda_i-1}, \quad
  c^i(z)=\sum_{s\in\bz}c^i_sz^{-s+\lambda_i},\quad
  (i=1,\cdots,N), \n
  &&
  \beta^a_r,\gamma^a_s,b^i_r,c^i_s|\mu,\lambda\rangle=0\;
  (r\geq 0,s\geq 1),
\eeqa
where some conditions will be imposed on $\mu_a$ and $\lambda_i$ later.
Then the $\WinfMN$ algebra with $C=1$ is realized as follows:
\beqa
  W(z^ne^{xD}A)
  &\!\!=\!\!&
  \oint\frac{dz}{2\pi i}
  \normal(\beta(z),b(z))z^ne^{xD}
  \left(
  \begin{array}{cc}
    A^{(0)}&A^{(+)} \\
    A^{(-)}&A^{(1)} \\
  \end{array}
  \right)
  {\gamma(z)\choose c(z)}\normal \n
  &\!\!=\!\!&
  \oint\frac{dz}{2\pi i}
  :(\beta(z),b(z))z^ne^{xD}
  \left(
  \begin{array}{cc}
    A^{(0)}&A^{(+)} \\
    A^{(-)}&A^{(1)} \\
  \end{array}
  \right)
  {\gamma(z)\choose c(z)}: \n
  &&
  +1\cdot
  \biggl(\sum_{a=1}^M\frac{e^{\mu_a x}-1}{e^x-1}A^{(0)}_{aa}
        -\sum_{i=1}^N\frac{e^{\lambda_i x}-1}{e^x-1}A^{(1)}_{ii}\biggr)
  \delta_{n0} \n
  &\!\!=\!\!&
  \sum_{a=1}^M\sum_{b=1}^M
  \sum_{{\scriptstyle r,s}\in\bz \atop
        {\scriptstyle r+s=n-\mu_a+\mu_b}}
  A^{(0)}_{ab}e^{x(\mu_b-s)}E^{ab}_0(r,s)
  +1\cdot\sum_{a=1}^M\frac{e^{\mu_a x}-1}{e^x-1}A^{(0)}_{aa}
  \delta_{n0} \n
  &&
  +\sum_{i=1}^N\sum_{j=1}^N
  \sum_{{\scriptstyle r,s}\in\bz \atop
        {\scriptstyle r+s=n-\lambda_i+\lambda_j}}
  A^{(1)}_{ij}e^{x(\lambda_j-s)}E^{ij}_1(r,s)
  -1\cdot\sum_{i=1}^N\frac{e^{\lambda_i x}-1}{e^x-1}A^{(1)}_{ii}
  \delta_{n0} \n
  &&
  +\sum_{a=1}^M\sum_{j=1}^N
  \sum_{{\scriptstyle r,s}\in\bz \atop
        {\scriptstyle r+s=n-\mu_a+\lambda_j}}
  A^{(+)}_{aj}e^{x(\lambda_j-s)}E^{aj}_+(r,s) \n
  &&
  +\sum_{i=1}^N\sum_{b=1}^M
  \sum_{{\scriptstyle r,s}\in\bz \atop
        {\scriptstyle r+s=n-\lambda_i+\mu_b}}
  A^{(-)}_{ib}e^{x(\mu_b-s)}E^{ib}_-(r,s).
  \Label{Wbcbg}
\eeqa
Here $E$'s are defined by
\beqa
  &&
  E^{ab}_0(r,s)=\;:\beta^a_r\gamma^b_s:,\quad
  E^{aj}_+(r,s)=\beta^a_rc^j_s, \n
  &&
  E^{ib}_-(r,s)=b^i_r\gamma^b_s,\qquad\;\;
  E^{ij}_1(r,s)=\;:b^i_rc^j_s:,
\eeqa
and they generate (super)matrix generalization of $\glinf$:
\beqa
  \Bigl[E^{ab}_0(r,s),E^{a'b'}_0(r',s')\Bigr]
  &\!\!=\!\!&
   \delta^{a'b}\delta_{r'+s,0}E^{ab'}_0(r,s')
  -\delta^{ab'}\delta_{r+s',0}E^{a'b}_0(r',s) \n
  &&
  -C\delta^{ab'}\delta^{a'b}\delta_{r+s',0}\delta_{r'+s,0}
  \Bigl(\theta(r\geq 0)-\theta(r'\geq 0)\Bigr),\n
  \Bigl[E^{ij}_1(r,s),E^{i'j'}_1(r',s')\Bigr]
  &\!\!=\!\!&
   \delta^{i'j}\delta_{r'+s,0}E^{ij'}_1(r,s')
  -\delta^{ij'}\delta_{r+s',0}E^{i'j}_1(r',s) \n
  &&
  +C\delta^{ij'}\delta^{i'j}\delta_{r+s',0}\delta_{r'+s,0}
  \Bigl(\theta(r\geq 0)-\theta(r'\geq 0)\Bigr),\n
  \Bigl\{E^{aj}_+(r,s),E^{ib}_-(r',s')\Bigr\}
  &\!\!=\!\!&
  \delta^{ij}\delta_{r'+s,0}E^{ab}_0(r,s')
  +\delta^{ab}\delta_{r+s',0}E^{ij}_1(r',s) \n
  &&
  -C\delta^{ab}\delta^{ij}\delta_{r+s',0}\delta_{r'+s,0}
  \Bigl(\theta(r\geq 0)-\theta(r'\geq 0)\Bigr),\n
  \Bigl[E^{ab}_0(r,s),E^{a'j}_+(r',s')\Bigr]
  &\!\!=\!\!&
   \delta^{a'b}\delta_{r'+s,0}E^{aj}_+(r,s'), \n
  \Bigl[E^{ab}_0(r,s),E^{ib'}_-(r',s')\Bigr]
  &\!\!=\!\!&
  -\delta^{ab'}\delta_{r+s',0}E^{ib}_-(r',s),\n
  \Bigl[E^{ij}_1(r,s),E^{aj'}_+(r',s')\Bigr]
  &\!\!=\!\!&
  -\delta^{ij'}\delta_{r+s',0}E^{aj}_+(r',s),\n
  \Bigl[E^{ij}_1(r,s),E^{i'b}_-(r',s')\Bigr]
  &\!\!=\!\!&
   \delta^{i'j}\delta_{r'+s,0}E^{ib}_-(r,s'),
\eeqa
where $C=1$ in this case and the other (anti-)commutation
relations vanish.

When $\mu_a$ and $\lambda_i$ satisfy the following condition,
\beqa
  \mu_a-\mu_b
  &\!\!=\!\!&
  0,\pm1, \n
  \lambda_i-\lambda_j
  &\!\!=\!\!&
  0,\pm1, \n
  \mu_a-\lambda_i
  &\!\!=\!\!&
  0,-1,
\eeqa
\eq{Wbcbg} implies that $|\mu,\lambda\rangle$ is the highest weight
state of $\WinfMN$,
\beqa
  W(z^nD^kA)|\mu,\lambda\rangle
  &\!\!=\!\!&
  0 \quad (n\geq 1,k\geq 0), \n
  W(D^kA^{(+)})|\mu,\lambda\rangle
  &\!\!=\!\!&
  0 \quad (k\geq 0), \n
  -W(e^{xD}E^{(0)}_{aa})|\mu,\lambda\rangle
  &\!\!=\!\!&
  -\frac{e^{\mu_a x}-1}{e^x-1}|\mu,\lambda\rangle,
  \;\mbox{\it i.e., }\;
  W(D^kE^{(0)}_{aa})|\mu,\lambda\rangle=
  -\Delta^{\mu_a}_k|\mu,\lambda\rangle, \n
  -W(e^{xD}E^{(1)}_{ii})|\mu,\lambda\rangle
  &\!\!=\!\!&
  \frac{e^{\lambda_i x}-1}{e^x-1}|\mu,\lambda\rangle,
  \;\mbox{\it i.e., }\;
  W(D^kE^{(1)}_{ii})|\mu,\lambda\rangle=
  \Delta^{\lambda_i}_k|\mu,\lambda\rangle.
\eeqa

The full character is defined by
\beq
  \mbox{ch}=\mbox{tr}\,\exp\sum_{k=0}^{\infty}\biggl(
  \sum_{a=1}^M{g'}^a_kW(D^kE^{(0)}_{aa})
  +\sum_{i=1}^Ng^i_kW(D^kE^{(1)}_{ii})
  \biggr),
\eeq
where the trace is taken over the irreducible representation space.
By taking a trace over whole Fock space,
we define
$S^{\mu_1,\cdots,\mu_M,\lambda_1,\cdots,\lambda_N}_{
    m'_1,\cdots,m'_M,m_1,\cdots,m_N}$ as follows:
\beqa
  &&
  \sum_{{{\scriptstyle m'_1,\cdots,m'_M}
        \atop {\scriptstyle m_1,\cdots,m_N}} \in\bz}
  S^{\mu_1,\cdots,\mu_M,\lambda_1,\cdots,\lambda_N}_{
     m'_1,\cdots,m'_M,m_1,\cdots,m_N}
  {t'_1}^{-m'_1}\cdots {t'_N}^{-m'_N}
  t_1^{-m_1}\cdots t_N^{-m_N} \n
  &\!\!=\!\!&
  e^{\sum_{k=0}^{\infty}\bigl(
    -\sum_{a=1}^M{g'}^i_k\Delta^{\mu_a}_k
    +\sum_{i=1}^Ng^i_k\Delta^{\lambda_i}_k\bigr)} \n
  &&\times
  \prod_{a=1}^M
  \prod_{r=1}^{\infty}\Bigl(1-t'_au_r(\mu_a)\Bigr)^{-1}
  \prod_{s=0}^{\infty}\Bigl(1-{t'}_a^{-1}v_s(\mu_a)\Bigr)^{-1}
  \biggl|_{g_k={g'}^a_k} \n
  &&\times
  \prod_{i=1}^N
  \prod_{r=1}^{\infty}\Bigl(1+t_iu_r(\lambda_i)\Bigr)
  \prod_{s=0}^{\infty}\Bigl(1+t_i^{-1}v_s(\lambda_i)\Bigr)
  \biggl|_{g_k=g^i_k},
\eeqa
where we have used
\beqa
  \Bigl[ W(D^kE^{(0)}_{aa}),\beta^b_{-r}\Bigr]
  =\delta_{ab}(\mu_a-r)^k\beta^b_{-r},
  &&
  \Bigl[ W(D^kE^{(0)}_{aa}),\gamma^b_{-s}\Bigr]
  =-\delta_{ab}(\mu_a+s)^k\gamma^b_{-s}, \n
  \Bigl[ W(D^kE^{(1)}_{ii}),b^j_{-r}\Bigr]
  =\delta_{ij}(\lambda_i-r)^kb^j_{-r},
  &&
  \Bigl[ W(D^kE^{(1)}_{ii}),c^j_{-s}\Bigr]
  =-\delta_{ij}(\lambda_i+s)^kc^j_{-s}.
\eeqa
$S^{\mu_1,\cdots,\mu_M,\lambda_1,\cdots,\lambda_N}_{
    m'_1,\cdots,m'_M,m_1,\cdots,m_N}$ can be expressed
in terms of $S^{\lambda;\epsilon}_m$,
\beq
  S^{\mu_1,\cdots,\mu_M,\lambda_1,\cdots,\lambda_N}_{
     m'_1,\cdots,m'_M,m_1,\cdots,m_N}
  =
  \prod_{a=1}^MS^{\mu_a;-1}_{m'_a}\biggl|_{g_k={g'}^a_k}
  \cdot
  \prod_{i=1}^NS^{\lambda_i;1}_{m_i}\biggl|_{g_k=g^i_k}.
\eeq
Since the sector of vanishing $\Uone$--charge in the Fock space
is again the irreducible representation space of $\WinfMN$,
the full character for the representation $|\mu,\lambda\rangle$ is
given by
\beq
  \mbox{ch }
  =
  \sum_{{{\scriptstyle m'_a,m_i}\in\bz
        \atop {\scriptstyle \sum_am'_a+\sum_im_i=0}}}
  S^{\mu_1,\cdots,\mu_M,\lambda_1,\cdots,\lambda_N}_{
     m'_1,\cdots,m'_M,m_1,\cdots,m_N}.
\eeq

Setting all ${g'}^a_k$ and $g^i_k$ to \eq{gk},
we obtain the specialized character.
For example, in the case of $M=0$, the specialized character is
essentially $\uone$
character times level $1$ $\widehat{su}(N)$ character\cite{rO}.
For $N=M=1$ and $\mu=\lambda$, the specialized character is
given by\footnote{
Eq.(79) in \cite{rO} can be expressed as
$ch^{W^{1,1}_{\infty}}_n(\theta,\tau)=
 \frac{1-q}{(1+zq^{n+\frac{1}{2}})(1+z^{-1}q^{-n+\frac{1}{2}})}
 \prod_{j=1}^{\infty}
 \frac{(1+zq^{j-\frac{1}{2}})(1+z^{-1}q^{j-\frac{1}{2}})}{(1-q^j)^2}$.
}
\beq
  \mbox{ch }
  =
  \frac{1}{1+q^{\frac{1}{2}}}
  \prod_{j=1}^{\infty}
  \biggl(\frac{1+q^{j-\frac{1}{2}}}{1-q^j}\biggr)^2.
\eeq

By interchanging $\beta\gamma$ with $bc$, we obtain the realization
with $C=-1$.
Although realizations with integer $C$ can be obtained by tensoring,
the character can not be derived by the method in this section.

\sectionnew{Quasifinite representation of $\Winf$}

We study the highest weight representation of $\Winf$.
The highest weight state $|\lambda\rangle$ is characterized by
\beqa
  W(z^nD^k)|\lambda\rangle
  &\!\!=\!\!&
  0 \quad (n\geq 1,k\geq 0), \n
  W(D^k)|\lambda\rangle
  &\!\!=\!\!&
  \Delta_k|\lambda\rangle \quad (k\geq 0),
\eeqa
where the weight $\Delta_k$ is some complex number.
It is convenient to introduce the generating function $\Delta(x)$ for
the weights $\Delta_k$:
\beq
  \Delta(x)=-\sum_{k=0}^{\infty}\Delta_k\frac{x^k}{k!},
\eeq
which we call the weight function. It is formally given as the
eigenvalue of the operator $-W(e^{xD})$:
\beq
  -W(e^{xD})|\lambda\rangle=\Delta(x)|\lambda\rangle.
\eeq
The Verma module is spanned by the state
\beq
  W(z^{-n_1}D^{k_1})\cdots W(z^{-n_m}D^{k_m})|\lambda\rangle.
  \Label{Wbasis}
\eeq
The energy level, which is the relative $L_0$ eigenvalue, of this
state is $\sum_{i=1}^mn_i$.
Reflecting the infinitely many generators, the Verma module has
infinitely many states at each level.
The irreducible representation space is obtained by subtracting
null states from the Verma module.
A null state is the state which can not be brought back
to $|\lambda\rangle$ by any successive actions of $\Winf$ generators.
Of course, in other words, a null state is the state
which has vanishing inner products with any states.

In the rest of this article, we will study the quasifinite
representations\cite{rKR}.
A representation is called {\it quasifinite} if there are only
a finite number of non-vanishing states at each energy level.
The representations obtained by free field realizations in the
previous section have this property, because there are only finite
number of oscillators at each energy level.
To achieve this, the weight function must be severely constrained.
We will show that if there are a finite number of states at level $1$,
then it is so at any level.

Let us assume that there are only a finite number of non-vanishing
states at level $n$.
This means that the following linear relation exists:
\beq
  W(z^{-n}f(D))|\lambda\rangle=\mbox{null},
\eeq
where $f$ is some polynomial. Acting $W(e^{x(D+n)})$ to this state,
we have
\beqan
  \mbox{null}
  &\!\!=\!\!&
  W(e^{x(D+n)})W(z^{-n}f(D))|\lambda\rangle \n
  &\!\!=\!\!&
  \Bigl[W(e^{x(D+n)}),W(z^{-n}f(D))\Bigr]|\lambda\rangle
  +\mbox{ null} \n
  &\!\!=\!\!&
  (1-e^{xn})W(z^{-n}e^{xD}f(D))|\lambda\rangle
  +\mbox{ null},
\eeqan
and thus the state $W(z^{-n}D^kf(D))|\lambda\rangle$ is also
null for all $k\geq 0$.
In other words, the set
\beq
  I_{-n}
  =
  \Bigl\{f(w)\in\bC[w]\Bigm|
  W(z^{-n}f(D))|\lambda\rangle=\mbox{null}\Bigr\}
\eeq
is an ideal in the polynomial ring $\bC[w]$. Since $\bC[w]$ is a
principal ideal domain, $I_{-n}$ is generated by a monic polynomial
$b_n(w)$, {\it i.e.} $I_{-n}=\{f(w)b_n(w)|f(w)\in\bC[w]\}$.
These polynomials $b_n(w)$ ($n=1,2,3,\cdots$) are called characteristic
polynomials for the quasifinite representation.

Let $f_n(w)$ be the minimal-degree monic polynomial satisfying
the following differential equation:
\beq
  f_n \left( \sfrac{d}{dx} \right)
  \sum_{j=0}^{n-1} e^{jx}
  \Bigl( (e^x-1)\Delta(x)+C \Bigr) = 0.
\eeq
Then the characteristic polynomials $b_n(w)$ are related to each
other as follows:\footnote{
We can also show that $b_{n+m}(w)$ divides $b_n(w-m)b_m(w)$.}
\beqan
  &\mbox{(\romannumeral1)}&
  \mbox{$b_n(w)$ divides both of $b_{n+m}(w+m)$ and $b_{n+m}(w)$
  ($m\geq 1$)}, \\
  &\mbox{(\romannumeral2)}&
  \mbox{ $f_n(w)$ divides $b_n(w)$}.
\eeqan
The $b_n(w)$'s are determined as the minimal-degree monic polynomials
satisfying both (\romannumeral1) and (\romannumeral2).
The property (\romannumeral1) is derived from the null state condition,
\beqan
  \mbox{null}
  &\!\!=\!\!&
  W(z^me^{x(D+n+m)})W(z^{-n-m}b_{n+m}(D))|\lambda\rangle \\
  &\!\!=\!\!&
  \Bigl[W(z^me^{x(D+n+m)}),W(z^{-n-m}b_{n+m}(D))\Bigr]|\lambda\rangle \\
  &\!\!=\!\!&
  W(z^{-n}(e^{xD}b_{n+m}(D)-e^{x(D+n+m)}b_{n+m}(D+m)))|\lambda\rangle.
\eeqan
The property (\romannumeral2) is derived from the following null state
condition,
\beqan
  0&\!\!=\!\!&
  W(z^ne^{x(D+n)})W(z^{-n}b_n(D))|\lambda\rangle \n
  &\!\!=\!\!&
  \Bigl[W(z^ne^{x(D+n)}),W(z^{-n}b_n(D))\Bigr]|\lambda\rangle \n
  &\!\!=\!\!&
  \biggl(W(e^{xD}b_n(D))-W(e^{x(D+n)}b_n(D+n))
  +C\sum_{j=1}^ne^{x(n-j)}b_n(n-j)\biggr)|\lambda\rangle \n
  &\!\!=\!\!&
  b_n\Bigl(\sfrac{d}{dx}\Bigr)
  \sum_{j=0}^{n-1} e^{jx}
  \Bigl( (e^x-1)\Delta(x)+C \Bigr)|\lambda\rangle.
\eeqan

The solution of these conditions are given by\cite{rKR,rAFMO3}
\beq
  b_n(w)={\rm lcm}(b(w),b(w-1),\cdots,b(w-n+1)),
  \Label{bn}
\eeq
where $b(w)=b_1(w)=f_1(w)$ is the minimal-degree monic polynomial
satisfying the differential equation,
\beq
  b \Bigl( \sfrac{d}{dx} \Bigr)
  \Bigl((e^x-1)\Delta(x)+C\Bigr) = 0.
  \Label{bd}
\eeq
Therefore, the necessary and sufficient condition for quasifiniteness is
that the weight function satisfies this type of differential
equation. Moreover it has been shown that
the finiteness at level 1 ({\it i.e.} existence of $b(w)$) implies
the finiteness at higher levels ({\it i.e.} existence of $b_n(w)$).

If we factorize the characteristic polynomial $b(w)$ as
\beq
  b(w)=\prod_{i=1}^K(w-\lambda_i)^{m_i},\quad
  (\lambda_i\neq\lambda_j),
  \Label{b}
\eeq
then the solution of \eq{bd} is given by
\beq
  \Delta(x)
  =
  \frac{\sum_{i=1}^Kp_i(x)e^{\lambda_ix}-C}{e^x-1},
\eeq
where $p_i(x)$ is a polynomial of degree $m_i-1$\footnote{
Since $b(w)$ is minimal-degree, $\deg p_i(x)$ is exactly $m_i-1$.}.
Since $\Delta(x)$ is regular at $x=0$ by definition, $p_i$'s satisfy
$\sum_{i=1}^Kp_i(0)=C$. Therefore $\Delta(x)$ has $\sum_{i=1}^K(m_i+1)$
parameters; $C$, $\lambda_i$ and coefficients in $p_i(x)$'s.
In contrast to the weight function for general (non-quasifinite)
representation, the weight function for quasifinite representation
has thus only finite parameters.
The representation realized by free field studied in section 3.1
has the weight function
$\Delta(x)=\epsilon\frac{e^{\lambda x}-1}{e^x-1}$,
which corresponds to the characteristic polynomial $b(w)=w-\lambda$.
We can explicitly check that $b_n(w)$ is given by \eq{bn}\cite{rM}.

Under the spectral flow \eq{sf}, the representation space as a set
is kept invariant. Furthermore the highest weight state with respect
to the original generators $W(\cdot)$ is also the highest weight state
with respect to the new generators $W'(\cdot)$.
On the other hand, the weight function $\Delta(x)$ and the
characteristic polynomial $b(w)$ are replaced by the new
ones\cite{rAFMO1}:
\beqa
  \Delta'(x)
  &\!\!=\!\!&
  e^{\lambda x}\Delta(x)+C\frac{e^{\lambda x}-1}{e^x-1}, \\
  b'(w)
  &\!\!=\!\!&
  b(w-\lambda).
\eeqa
This implies that the spectral flow transforms $\lambda_i$ in \eq{b}
into $\lambda_i+\lambda$.

To study the structure of null states,
let us introduce the inner product as
\beqa
  &&
  \langle\lambda|\lambda\rangle=1, \n
  &&
  \Bigl(\langle\lambda|W\Bigr)|\lambda\rangle=
  \langle\lambda|\Bigl(W|\lambda\rangle\Bigr)=
  \langle\lambda|W|\lambda\rangle,
\eeqa
and the corresponding bra state $\langle\lambda|$ as
\beqa
  \langle\lambda|W(z^nD^k)
  &\!\!=\!\!&
  0 \quad (n\leq -1,k\geq 0), \n
  \langle\lambda|W(D^k)
  &\!\!=\!\!&
  \Delta_k\langle\lambda| \quad (k\geq 0).
\eeqa
Then the quasifinite condition for bra states is
\beq
  \langle\lambda|W(z^nb_n(D+n)D^k)=\mbox{null}\quad
  (n\geq 1,k\geq 0).
\eeq
These are consistent with the hermitian conjugation \eq{dagger}
when $\Delta_k\in\bR$ (or $\Delta_k\in\bC$ if $\dagger$ is modified,
$(aA+bB)^{\dagger}=aA^{\dagger}+bB^{\dagger}$).

Unitary representation was studied in \cite{rKR}.
The necessary and sufficient condition for unitary representation
is that $C$ is a non-negative integer and the weight function is
\beq
  \Delta(x)=\sum_{i=1}^C\frac{e^{\lambda_ix}-1}{e^x-1},\quad
  \lambda_i\in\bR.
\eeq
This weight function corresponds to the characteristic polynomial
$b(w)=\prod'_i(w-\lambda_i)$ where the product is taken over
different $\lambda_i$.
We remark that all the unitary representations can be realized
by tensoring $C$ pairs of $bc$ ghosts, \eq{tensor}.

Quasifinite representations of $\WinfMN$ and subalgebras
can be treated similarly (For $M=N=1$, see \cite{rAFMO2},
and for $\Winfp$ see \cite{rAFMO4}).

\sectionnew{Kac determinant and full character formulae of $\Winf$}

In this section we compute the Kac determinant for some representations,
on the basis of which we derive the analytic form of the Kac
determinant and full character formulae.

\subsection{The Verma module}

Let us study the quasifinite representation of $\Winf$ with
central charge $C$ and the weight function $\Delta(x)$.
Characteristic polynomials $b_n(w)$ are determined from $\Delta(x)$.
Since there are linear relations $W(z^{-n}D^kb_n(D))|\lambda\rangle=$
null, only $\deg b_n(w)$ states are independent in the states
$\{W(z^{-n}D^k)|\lambda\rangle\}$. We may take independent states as
follows:
\beq
  W(z^{-n}D^k)|\lambda\rangle \quad
  (k=0,1,\cdots,\deg b_n(w)-1).
\eeq
The Verma module for the quasifinite representation is defined as
the space spanned by these generators. Therefore the specialized
character formula for the Verma module is
\beq
  \mbox{tr}\,q^{L_0}
  =
  q^{-\Delta_1}\prod_{n=1}^{\infty}\frac{1}{(1-q^n)^{\deg b_n(w)}}.
  \Label{Verma}
\eeq
For example, for $b(w)=w-\lambda$, we have
\beq
  \mbox{tr}\,q^{L_0+\Delta_1^{\lambda}}
  =\chi(q)=\prod_{n=1}^{\infty}\frac{1}{(1-q^n)^n}.
  \Label{chi}
\eeq
This $\chi(q)$ has a close relationship with the partition function
of three--dimensional free field theory (see \cite{rAFMO1}).

\subsection{Determinant formulae at lower levels}

In this subsection we will present explicit computation of
the Kac determinant for quasifinite representations\cite{rAFMO1}.
First let us consider the representation with
$\Delta(x)=C\frac{e^{\lambda x}-1}{e^x-1}$.
The characteristic polynomial is $b(w)=w-\lambda$ ($b(w)=1$ for
$C=\lambda=0$).
For the first three levels, the relevant ket states are,
\beqa
  \mbox{Level 1}
  &\quad&
  W(z^{-1})|\lambda\rangle, \n
  \mbox{Level 2}
  &\quad&
  W(z^{-2})|\lambda\rangle,W(z^{-1})^2|\lambda\rangle,
  W(z^{-2}D)|\lambda\rangle, \n
  \mbox{Level 3}
  &\quad&
  W(z^{-3})|\lambda\rangle,W(z^{-1})W(z^{-2})|\lambda\rangle,
  W(z^{-1})^3|\lambda\rangle, \n
  &\quad&
  W(z^{-3}D)|\lambda\rangle,
  W(z^{-1})W(z^{-2}D)|\lambda\rangle,
  W(z^{-3}D^2)|\lambda\rangle.
\eeqa
Corresponding bra states may be given by changing $z^{-n}$ into $z^n$.
The number of relevant states grows as,
\beq
  \chi(q)
  =
  1+q+3q^2+6q^3+13q^4+24q^5+48q^6+86q^7+160q^8+282q^9+500q^{10}+\cdots.
  \Label{chi2}
\eeq
Inner product matrices are straightforwardly calculated;
for example, at level 2,
$$
  \left(
  \begin{array}{ccc}
    2C&0&(2\lambda+1)C \\
    0&2C^2&-C \\
    (2\lambda-3)C&-C&(2\lambda^2-2\lambda-1)C
  \end{array}
  \right).
$$
The determinant for this matrix is $2C^3(C-1)$.
We computed the Kac determinant up to level 8 by using
computer\cite{rAFMO1}:
\beqan
  \det[1]&\!\!\propto\!\!& C, \n
  \det[2]&\!\!\propto\!\!& C^3(C-1), \n
  \det[3]&\!\!\propto\!\!& C^6(C-1)^3(C-2), \n
  \det[4]&\!\!\propto\!\!& (C+1)C^{13}(C-1)^8(C-2)^3(C-3), \n
  \det[5]&\!\!\propto\!\!& (C+1)^3C^{24}(C-1)^{17}(C-2)^8(C-3)^3(C-4), \n
  \det[6]&\!\!\propto\!\!& (C+1)^{10}C^{48}(C-1)^{37}(C-2)^{19}(C-3)^{8}
  (C-4)^3(C-5), \n
  \det[7]&\!\!\propto\!\!& (C+1)^{23}C^{86}(C-1)^{71}(C-2)^{41}(C-3)^{19}
  (C-4)^{8}(C-5)^{3}(C-6), \n
  \det[8]&\!\!\propto\!\!& (C+1)^{54}C^{161}(C-1)^{138}(C-2)^{85}
  (C-3)^{43}(C-4)^{19}(C-5)^{8}(C-6)^{3}(C-7).
\eeqan
We remark that $\lambda$-dependence disappears
due to nontrivial cancellations.
This is explained by the spectral flow \cite{rAFMO1}.
We computed also the corank of the inner product matrix:
\beqan
  \mbox{cor}[n]&\!\!=\!\!& \det[n], \quad (n=1,\cdots,7), \n
  \mbox{cor}[8]&\!\!=\!\!& (C+1)^{54}C^{160}(C-1)^{138}(C-2)^{85}
  (C-3)^{43}(C-4)^{19}(C-5)^{8}(C-6)^{3}(C-7),
\eeqan
where the exponent stands for a corank, {\it i.e.} a number
of null states.
Subtracting this number from \eq{chi2}, we get the specialized
characters at lower levels.
We will present the determinant and full character formulae
in section 5.3.

Next we take the representation with
\beq
  \Delta(x)=\sum_{i=1}^KC_i\frac{e^{\lambda_ix}-1}{e^x-1}, \quad
  C=\sum_{i=1}^KC_i,
\eeq
where $\lambda_i$'s are all different numbers.
The characteristic polynomial is given by
\beq
  b(w)=\prod_{i=1}^K(w-\lambda_i)
\eeq
(if $C_i=\lambda_i=0$, then the factor $w-\lambda_i$ in $b(w)$ should
be omitted).
Assuming that the difference of any two $\lambda_i$'s is not an integer,
we computed the Kac determinants at lower levels for $K=1,\cdots,5$,
and they are given in appendix A\cite{rAFMO1}.
The determinant formula may be written in the following form:
\beq
  \det[n]\propto\prod_iA_n(C_i)\prod_{i<j}B_n(\lambda_i-\lambda_j).
  \Label{det}
\eeq
$\lambda_i$-dependence appears only through their differences
due to the spectral flow symmetry \cite{rAFMO1,rAFMO3}.
The functions $A_n$ and $B_n$ have zero only when $C_i$ or
$\lambda_i-\lambda_j$ is integer.

We will give an analytic expression for $B_n(\lambda)$.
Let us consider the case when one pair $\lambda_i-\lambda_j$ is
an integer $\ell$. In this case the characteristic polynomial $b_n(w)$
may have degree less than $n\deg b(w)$.
The weight function becomes
\beq
  \Delta'(x;\ell)=\Delta(x)\biggl|_{\lambda_i-\lambda_j=\ell},
\eeq
and we denote the corresponding characteristic polynomial
as $b'(w;\ell)$ and $b'_n(w;\ell)$, and the discrepancy
of degree as
\beq
  d(n;\ell)=n\deg b'(w;\ell)-\deg b'_n(w;\ell).
\eeq
Then $B_n(\lambda)$ is
\beq
  B_n(\lambda)
  =
  \prod_{\ell\in\bz}(\lambda-\ell)^{\beta_n(\ell)},
\eeq
where $\beta_n(\ell)$ is defined by
\beq
  \sum_{n=0}^{\infty}\beta_n(\ell)q^n
  =
  2t\frac{d}{dt}
  \prod_{n=1}^{\infty}
  \frac{(1-q^n)^{d(n;\ell)}}{(1-tq^n)^{d(n;\ell)}}
  \biggl|_{t=1} \cdot\,
  \chi(q)^{\deg b'(w;\ell)}.
\eeq
Its proof can be found in \cite{rAFMO3}.

We remark that the determinant formula for the representation
with $\Delta'(x;\ell)$ is not given by \eq{det},
because \eq{det} is the determinant for the inner product matrix
of size given by \eq{Verma} with $b_n(w)$ not $b'_n(w,\ell)$.
Such a determinant has mixed $C_i$ factors, {\it e.g.} $C_i+C_j+1$.
%
%
We here give some examples in the $\lambda_i-\lambda_{i+1}=1$ case.
When $K=2$, the determinant for the first five levels are
\beqan
   \det[1]&\!\!\propto\!\!& C_1 C_2, \n
   \det[2]&\!\!\propto\!\!& (C_1+C_2+1)
      \prod_{i=1,2} C_i^3(C_i-1),\n
   \det[3]&\!\!\propto\!\!& (C_1+C_2+1)^4
      \prod_{i=1,2} C_i^{8} (C_i-1)^3 (C_i-2),\n
   \det[4]&\!\!\propto\!\!& (C_1+C_2+1)^{13}(C_1+C_2)
      \prod_{i=1,2} C_i^{20} (C_i-1)^9 (C_i-2)^3 (C_i-3),\n
   \det[5]&\!\!\propto\!\!& (C_1+C_2+1)^{34}(C_1+C_2)^4 
      \prod_{i=1,2} C_i^{46} (C_i-1)^{22} (C_i-2)^{9} (C_i-3)^3 (C_i-4),
\eeqan
and when $K=3$, that of the first three levels are
\beqan
   \det[1]&\!\!\propto\!\!& \prod_{i=1,2,3} C_i, \n
   \det[2]&\!\!\propto\!\!&   (C_1+C_2+1) (C_2+C_3+1)
      \prod_{i=1,2,3} C_i^4 (C_i-1), \n
   \det[3]&\!\!\propto\!\!& (C_1+C_2+1)^4 (C_2+C_3+1)^4 (C_1+C_2+C_3+2)
      \prod_{i=1,2,3}  C_i^{13} (C_i-1)^4 (C_i-2).
\eeqan

\subsection{$\Delta(x)=C\frac{e^{\lambda x}-1}{e^x-1}$ case}

In this subsection we study the representation with
$\Delta(x)=C\frac{e^{\lambda x}-1}{e^x-1}$\cite{rAFMO3}.
The corresponding characteristic polynomial is $b(w)=w-\lambda$
($b(w)=1$ for $C=\lambda=0$, but in this case non-vanishing
states are $|\lambda\rangle$ only. So we do not care about
this case).
We have already known the full character formula for $C=\pm 1$.
The numbers of states, eqs.\,(\ref{S1b},\ref{Sm1b}), are less than
that of the Verma module \eq{chi}.
The determinant formula at lower levels given in the previous
section suggests that additional
null states appear only when $C$ is an integer.
This is indeed the case, and we will derive the analytic form
of the determinant and full character formulae.

As we have shown, the basis \eq{Wbasis} is not a good basis because
of \eq{WDkW}. The construction of the diagonal basis becomes possible
if we view the $\Winf$ algebra from the equivalent $\glinf$ algebra,
which is implicitly used in section 3.1.
It was proved that the quasifinite representations of those algebras
coincide\cite{rKR}. The $\glinf$ algebra is defined by \eq{glinf}
and the relation with $\Winf$
is\footnote{
This relation should be modified for different $b(w)$.
When $b(w)=0$ has multiple roots, for example $b(w)=(w-\lambda)^m$,
$\glinf$ also need to be modified\cite{rKR,rAFMO3}.}
\beq
  W(z^ne^{xD})
  =
  \sum_{{\scriptstyle r,s}\in\bz \atop {\scriptstyle r+s=n}}
  e^{x(\lambda-s)}E(r,s)
  -C\frac{e^{\lambda x}-1}{e^x-1}\delta_{n0}.
\eeq
The highest weight state of $\glinf$ is defined by
\beqa
  E(r,s)|\lambda\rangle
  &\!\!=\!\!&
  0 \quad (r+s>0),\n
  E(r,-r)|\lambda\rangle
  &\!\!=\!\!&
  q_r|\lambda\rangle \quad (r\in\bZ).
\eeqa
The quasifiniteness of the representation is achieved only when
finite number of $h_r=q_r-q_{r-1}+C\delta_{r0}$ are
non-vanishing\cite{rKR}.
In this case the following $E(r,s)$ annihilates the highest weight
state\cite{rAFMO3}:
\beq
  E(r,s)|\lambda\rangle=0\quad (r\geq 0,s\geq 1).
\eeq

The generator $E(r,s)$ is already diagonal with respect to the
action of the Cartan elements,
\beq
  \Bigl[W(D^k),E(r,s)\Bigr]
  =
  \Bigl((\lambda+r)^k-(\lambda-s)^k\Bigr)E(r,s).
  \Label{WDkE}
\eeq
Therefore the state
\beq
  E(-r_1,-s_1)\cdots E(-r_n,-s_n)|\lambda\rangle,\quad
  (r_a\geq 1,s_a\geq 0),
  \Label{Ebasis}
\eeq
is the simultaneous eigenstate of $W(D^k)$ with the eigenvalues
\beq
  \Delta^{\lambda}_k+\sum_{a=1}^n
  \Bigl((\lambda-r_a)^k-(\lambda+s_a)^k\Bigr).
\eeq
The representation space is decomposed into the eigenspace with
above eigenvalues. So we need to consider only this subspace,
which is spanned by
\beq
  \prod_{a=1}^nE(-r_a,-s_{\sigma(a)})|\lambda\rangle,\quad
  (r_a\geq 1,s_a\geq 0),
  \Label{Ebasis2}
\eeq
where $\sigma$ is a permutation of $n$ objects.
The number of these states is equal to the number of onto-map
from $I=\{r_1,\cdots,r_n\}$ to $J=\{s_1,\cdots,s_n\}$.

We calculated the inner product matrix of these state\cite{rAFMO3}.
By symmetrizing the indices of \eq{Ebasis2} according to the Young
diagram with $n$ boxes, this matrix can be block-diagonalized.
Each block can be further diagonalized.
In fact, when $r_a$'s and $s_a$'s
are all different respectively, we obtained an explicit form
$|Y;\alpha,\beta\rangle$ (see \cite{rAFMO3} for details).
In general, by taking appropriate linear combination of \eq{Ebasis2},
an orthogonal basis $|Y;\alpha,\beta\rangle$ ($\alpha=1,\cdots,d_Y^I$;
$\beta=1,\cdots,d_Y^J$) is obtained:
\beq
  \langle Y;\alpha,\beta|Y';\alpha',\beta'\rangle
  =
  \delta_{YY'}\delta_{\alpha\alpha'}\delta_{\beta\beta'}
  \frac{\sqrt{d_Y^Id_Y^J}}{n!}\prod_{b\in Y}(C-C_b).
  \Label{Ynorm}
\eeq
Here, to each box $b$ in the Young diagram $Y$,
we assign a number $C_b$ as
\beq
  \begin{tabular}{|c|c|c|c|c}\hline
    $0$&$1$&$2$&$\;3\;$&$\cdots$\\ \hline
    $-1$&$0$&$1$&$2$&$\cdots$\\ \hline
    $-2$&$-1$&$0$&$1$&$\cdots$\\ \hline
    $-3$&$-2$&$-1$&$0$&$\cdots$\\ \hline
    $\vdots$&$\vdots$&$\vdots$&$\vdots$&$\ddots$
  \end{tabular}.
  \Label{Cb}
\eeq
$d_Y^I$ is the number of assignment of $r_a$ to each box in the
Young diagram $Y$ with $n$ boxes such that $r_a$'s are non-decreasing
from left to right, and increasing from top to bottom.
When $r_a$'s are all different, $d_Y^I$ is equal to $d_Y$, the
dimension of irreducible representation $Y$ of permutation group
$\Sn$.

The determinant formula given in section 5.2 is reproduced from
above results. For example, at level 4,
\beqan
  \det[4]
  &\!\!\propto\!\!&
  (C+1)C^{13}(C-1)^8(C-2)^3(C-3) \\
  &\!\!=\!\!&
  C\times C\times C\times C\times C(C-1)\times C(C-1) \\
  &&
  \times C(C-1)\cdot C(C+1)\times C(C-1)\times C(C-1)
  \times C(C-1)(C-2) \\
  &&
  \times C(C-1)(C-2)\times C(C-1)(C-2)(C-3),
\eeqan
where each factor comes from
\beqan
  (I,J)
  &\!\!=\!\!&
  (\{4\},\{0\}),(\{3\},\{1\}),(\{2\},\{2\}),(\{1\},\{3\}),
  (\{3,1\},\{0,0\}),(\{2,2\},\{0,0\}), \\
  &&
  (\{2,1\},\{1,0\}),(\{1,1\},\{2,0\}),(\{1,1\},\{1,1\}),
  (\{2,1,1\},\{0,0,0\}), \\
  &&
  (\{1,1,1\},\{1,0,0\}),(\{1,1,1,1\},\{0,0,0,0\}).
\eeqan

As a simple corollary of the inner product formula,
we may derive the condition for the unitarity.
The positivity of the representation space may be rephrased as
the positivity of the right hand side of \eq{Ynorm} for any $Y$.
{}From \eq{Cb}, we can immediately prove that
this condition is achieved only when $C$ is non-negative integer.

The full character is defined by \eq{ch}.
{}From above determinant formula, when $C$ is not an integer,
there are no null states aside from those coming from characteristic
polynomials. Therefore combining \eqs{WDkE}{Ebasis}
we get the full character formula for non-integer $C$,
\beq
  \mbox{ch}
  =
  e^{C\sum_{k=0}^{\infty}g_k\Delta^{\lambda}_k}
  \prod_{r=1}^{\infty}\prod_{s=0}^{\infty}
  \frac{1}{1-u_r(\lambda)v_s(\lambda)},
\eeq
where $\Delta^{\lambda}_k$, $u_r(\lambda)$ and $v_s(\lambda)$ are
defined by \eq{dl},\eqn{uv} respectively.

If we expand this product as
$$
  \sum_{n=0}^{\infty}
  \sum_{{\scriptstyle I,J}\atop{\scriptstyle |I|=|J|=n}}
  N(I,J)\prod_{r\in I}u_r(\lambda)\prod_{s\in J}v_s(\lambda),
$$
then $N(I,J)$ gives the number of the states of the form \eq{Ebasis2}.
We need to go further to classify those states after the Young diagram.
The following result gives such classification (see appendix B),
\beq
  \prod_{r=1}^{\infty}\prod_{s=0}^{\infty}
  \frac{1}{1-u_r(\lambda)v_s(\lambda)}
  =
  \sum_{Y}\tau_Y(x(\lambda))\tau_Y(y(\lambda)),
  \Label{tautau}
\eeq
where the summation is taken over all Young diagrams,
and $\tau_{Y}$ is the character of irreducible representation $Y$
of $\glinf$, and the parameters $x$ and $y$ are
the Miwa variables for $u$ and $v$ defined by \eq{xy}.
If we expand each factor in the summation, we can get
the degeneracy with respect to each Young diagram $Y$,
and the eigenvalues.
The coefficient of $\prod_{r\in I}u_r(\lambda)$ in $\tau_Y(x(\lambda))$
is $d_Y^I$, and $N(I,J)=\sum_Yd_Y^Id_Y^J$.

Combining these Young diagram classification \eqs{Ynorm}{tautau},
we get the full character formula with integer $C$ \cite{rAFMO3},
\beqa
  \mbox{ch}_{C=n}
  &\!\!=\!\!&
  e^{n\sum_{k=0}^{\infty}g_k\Delta^{\lambda}_k}
  \sum_{{\scriptstyle Y}\atop{\scriptstyle wd(Y)\leq n}}
  \tau_Y(x(\lambda))\tau_Y(y(\lambda)),
  \Label{chn} \\
  \mbox{ch}_{C=-n}
  &\!\!=\!\!&
  e^{-n\sum_{k=0}^{\infty}g_k\Delta^{\lambda}_k}
  \sum_{{\scriptstyle Y}\atop{\scriptstyle ht(Y)\leq n}}
  \tau_Y(x(\lambda))\tau_Y(y(\lambda)),
  \Label{chmn}
\eeqa
where $n$ is an non-negative integer, and
$wd(Y)$ ($ht(Y)$) stands for the number of columns (rows) of $Y$.
The full characters obtained in section 3.1 agree with this result.

By setting $g_k$ to \eq{gk}, we obtain the specialized character.
In this case the Schur polynomial is expressed as
\beqa
  \tau_Y(x(\lambda))
  &\!\!=\!\!&
  q^{(1-\lambda)\sum_{j=1}^njm_j+\sum_{j=1}^n\frac{1}{2}j(j-1)m_j}
  \prod_{k=1}^nF_k(q;m_1,\cdots,m_k), \n
  \tau_Y(y(\lambda))
  &\!\!=\!\!&
  q^{\lambda\sum_{j=1}^njm_j+\sum_{j=1}^n\frac{1}{2}j(j-1)m_j}
  \prod_{k=1}^nF_k(q;m_1,\cdots,m_k), \n
  \tau_{{}^tY}(x(\lambda))
  &\!\!=\!\!&
  q^{(1-\lambda)\sum_{j=1}^njm_j+\frac{1}{2}\sum_{j=1}^n
  \bigl(\sum_{s=j}^n m_s\bigr)\bigl(\sum_{s=j}^n m_s-1\bigr)}
  \prod_{k=1}^nF_k(q;m_1,\cdots,m_k), \n
  \tau_{{}^tY}(y(\lambda))
  &\!\!=\!\!&
  q^{\lambda\sum_{j=1}^njm_j+\frac{1}{2}\sum_{j=1}^n
  \bigl(\sum_{s=j}^n m_s\bigr)\bigl(\sum_{s=j}^n m_s-1\bigr)}
  \prod_{k=1}^nF_k(q;m_1,\cdots,m_k),
\eeqa
where $Y=(m_1+\cdots+m_n,m_2+\cdots+m_n,\cdots,m_n)$,
and ${}^tY$ is the transpose of the Young diagram $Y$,
and $F_k(q;m_1,\cdots,m_k)$ is
\beq
  F_k(q;m_1,\cdots,m_k)
  =
  \prod_{j=1}^{m_k}\prod_{s=0}^{k-1}
  \Bigl(1-q^{\sum_{t=1}^s m_{k-t}+s+j}\Bigr)^{-1}.
\eeq
This expression is obtained from \eq{qtaur} by setting
$f_i-f_{i+1}=m_i$ (or \eq{qtauc} by setting $g_i-g_{i+1}=m_i$).
The full characters \eqs{chn}{chmn} reduce to the specialized
characters,
\beqa
  \mbox{ch}_{C=n}
  &\!\!=\!\!&
  q^{\frac{1}{2}\lambda(\lambda-1)n}
  \sum_{m_1=0}^{\infty}\cdots\sum_{m_n=0}^{\infty}
  q^{\sum_{j=1}^{n}\bigl(\sum_{s=j}^n m_s\bigr)^2}
  \prod_{k=1}^nF_k(q;m_1,\cdots,m_k)^2,
  \Label{qchn} \\
  \mbox{ch}_{C=-n}
  &\!\!=\!\!&
  q^{-\frac{1}{2}\lambda(\lambda-1)n}
  \sum_{m_1=0}^{\infty}\cdots\sum_{m_n=0}^{\infty}
  q^{\sum_{j=1}^n j^2 m_j}
  \prod_{k=1}^nF_k(q;m_1,\cdots,m_k)^2.
  \Label{qchmn}
\eeqa
As we will show in the next subsection, \eq{qchn} can be
rewritten in a product form:
\beq
  \mbox{ch}_{C=n}
  =
  q^{\frac{1}{2}\lambda(\lambda-1)n}
  \prod_{j=1}^{\infty}\prod_{k=1}^n\frac{1}{1-q^{j+k-1}}.
  \Label{qchnp}
\eeq
This character is consistent with the conjecture that the
representation space is spanned by $W(z^{-j}D^{k-1})$ with
$1\leq k\leq n$
(of course with $j\geq 1$, $1\leq k\leq j$)\cite{rAFMO1}.

\subsection{Other cases}

In ref.\cite{rFKRW}, the quasifinite representation with
the weight function,
\beq
  \Delta(x)
  =
  \sum_{i=1}^N\frac{e^{\lambda'_ix}-1}{e^x-1},\quad
  C=N,
\eeq
was studied. This representation is realized by $bc$ ghost,
\eq{tensor}.
We review their results.
Let us break the set $\{\lambda'_1,\cdots,\lambda'_N\}$ in the
following way:
\beqa
  &&
  \{\lambda'_1,\cdots,\lambda'_N\}
  =
  S_1\cup\cdots\cup S_m, \n
  &&
  S_i=\{\lambda_i+k^{(i)}_1,\cdots,\lambda_i+k^{(i)}_{n_i}\},\quad
  \lambda_i-\lambda_j\not\in\bZ,\quad
  k^{(i)}_1\geq\cdots\geq k^{(i)}_{n_i}\in\bZ.
\eeqa
Then the representation  for the weight function
$\Delta(x)$ is a direct product of the  representations  for
$\Delta_i(x)=
\sum_{j=1}^{n_i}\frac{e^{(\lambda_i+k^{(i)}_j)x}-1}{e^x-1}$ \cite{rKR}.
Therefore, the character is factorized as
\beq
  \mbox{ch}=\prod_{i=1}^m\mbox{ch}_i,
\eeq
where ch${}_i$ is the character of the representation for
$\Delta_i(x)$, and
we need to consider only the quasifinite representations with
\beq
  \Delta(x)
  =
  \sum_{i=1}^n\frac{e^{(\lambda+k_i)x}-1}{e^x-1},\quad
  C=n,\quad
  k_1\geq\cdots\geq k_n\geq 0\in\bZ.
  \Label{dki}
\eeq
The full characters for these representations are given by \cite{rFKRW}
\beq
  \mbox{ch}
  =
  \det\Bigl(S_{\lambda+k_i-i+j}\Bigr)_{1\leq i,j\leq n},
  \Label{detS}
\eeq
where $S_{\lambda}$ is defined in section 3.1.
By setting $g_k$ to \eq{gk} and using \eq{S1b},
\eq{detS} reduces to the specialized character,
\beq
  \mbox{ch}
  =
  q^{\sum_{i=1}^n\frac{1}{2}(\lambda+k_i)(\lambda+k_i-1)}
  \prod_{j=1}^{\infty}\frac{1}{(1-q^j)^n}
  \prod_{1\leq i<j\leq n}(1-q^{k_i-k_j-i+j}).
  \Label{qchp}
\eeq

For $C=n>0$, the weight function
$\Delta(x)=C\frac{e^{\lambda x}-1}{e^x-1}$ studied in the previous
section is a special case of \eq{dki}; $k_1=\cdots=k_n=0$.
So the full character \eq{chn} must be obtained from \eq{detS}.
In fact we can show that the full character \eq{detS} is expressed
as a summation over all Young diagrams:
\beqa
  \mbox{ch}
  &\!\!=\!\!&
  \det\Bigl(S_{\lambda+k_i-i+j}\Bigr)_{1\leq i,j\leq n} \n
  &\!\!=\!\!&
  \det\biggl((-1)^{j-i+k_i}
  e^{\sum_{k=0}^{\infty}g_k\Delta^{\lambda}_k}
  \sum_{m\in\bz}P_{m-n-1+i-k_i}(-x(\lambda))
                P_{m-n-1+j}(-y(\lambda))
  \biggr)_{1\leq i,j\leq n} \n
  &\!\!=\!\!&
  (-1)^{|Y|}
  e^{n\sum_{k=0}^{\infty}g_k\Delta^{\lambda}_k}
  \sum_{h_1\geq\cdots\geq h_n\geq 0}
  \det\Bigl(P_{h_j-k_i+i-j}(-x(\lambda)\Bigr)_{1\leq i,j\leq n}
  \det\Bigl(P_{g_i-i+j}(-y(\lambda)\Bigr)_{1\leq i,j\leq n} \n
  &\!\!=\!\!&
  e^{n\sum_{k=0}^{\infty}g_k\Delta^{\lambda}_k}
  \sum_{{\scriptstyle Y}\atop{\scriptstyle wd(Y)\leq n}}
  \tau_{Y/Y_k}(x(\lambda))\tau_Y(y(\lambda)),
\eeqa
where $Y_k$ is a Young diagram with ${}^tY_k=(k_1,\cdots,k_n)$,
and $\tau_{Y/Y_k}$ is a skew S--function (see appendix B).
Here we have used the determinant formula for the product of
non-square matrices,
\beq
  \det\biggl(\sum_{m=1}^Na_{im}b_{jm}\biggr)_{1\leq i,j\leq n}
  =
  \sum_{1\leq m_1<\cdots<m_n\leq N}
  \det\Bigl(a_{im_j}\Bigr)_{1\leq i,j\leq n}
  \det\Bigl(b_{im_j}\Bigr)_{1\leq i,j\leq n}.
\eeq
For $k_1=\cdots=k_n=0$, $\tau_{Y/Y_k}$ reduces to $\tau_Y$.
So we establish the equivalence of \eq{chn} and \eq{detS} in
this case. Eq.\,(\ref{qchnp}) is obtained from \eq{qchp}.

Similarly we can rewrite the full character \eq{chmn} as
\beq
  \mbox{ch}_{C=-n}
  =
  \det\Bigl(S^{\lambda;-1}_{j-i}\Bigr)_{1\leq i,j\leq n}.
\eeq

\subsection{Differential equation for full characters}

Finally we comment on the differential equation of the full character.
{}From \eq{Sgen}, $S^{\lambda;\epsilon}_m$ as a function of $x$ and
$y$ satisfies the differential equation,
\beq
  \frac{\partial}{\partial x_{\ell}}S^{\lambda;\epsilon}_m
  =
  (-\epsilon)^{\ell+1}S^{\lambda;\epsilon}_{m+\ell},\quad
  \frac{\partial}{\partial y_{\ell}}S^{\lambda;\epsilon}_m
  =
  (-\epsilon)^{\ell+1}S^{\lambda;\epsilon}_{m-\ell}.
\eeq
Thus the full character \eq{detS} satisfies the following
differential equation,
\beqa
  \frac{\partial}{\partial x_{\ell}}
  S^{\lambda}_{\{k_1,\cdots,k_n\}}
  &\!\!=\!\!&
  (-1)^{\ell+1}\sum_{i=1}^n
  S^{\lambda}_{\{k_1,\cdots,k_i+\ell,\cdots,k_n\}},\n
  \frac{\partial}{\partial y_{\ell}}
  S^{\lambda}_{\{k_1,\cdots,k_n\}}
  &\!\!=\!\!&
  (-1)^{\ell+1}\sum_{i=1}^n
  S^{\lambda}_{\{k_1,\cdots,k_i-\ell,\cdots,k_n\}},
\eeqa
where $S^{\lambda}_{\{k_1,\cdots,k_n\}}=
\det(S_{\lambda+k_i-i+j})_{1\leq i,j\leq n}$.

\vskip 5mm
\noindent{\bf Acknowledgments:}
S.O. would like to thank members of YITP for their hospitality
and the organizers of the meeting for giving him the opportunity
to present our results.
This work is supported in part by Grant-in-Aid for Scientific
Research from Ministry of Science and Culture.

\section* {Appendix A: Determinant formulae at lower degrees}
\setcounter{section}{1}
\renewcommand{\thesection}{\Alph{section}}
\renewcommand{\theequation}{\Alph{section}.\arabic{equation}}
\setcounter{equation}{0}

In this appendix, we give the explicit form of
the functions $A_n(C)$ and $B_n(\lambda)$
defined in \eq{det}\cite{rAFMO1}.
We can parametrize those functions in the form,
$$
  A_n(C)=\prod_{\ell\in\bz}(C-\ell)^{\alpha(\ell)},
  \qquad
  B_n(\lambda)=\prod_{\ell\in\bz}(\lambda-\ell)^{\beta(\ell)}
$$
We make tables for the index $\alpha(\ell)$ and $\beta(\ell)$.
We note that $\beta(\ell)=\beta(-\ell)$.  Hence we will write
them only for $\ell\geq 0$.

\noindent \underline{$K=1$}:\hskip 4mm $B_n=1$ due to the
spectral flow symmetry \cite{rAFMO1}.
\begin{center}
\begin{tabular}{|c||c|c|c|c|c|c|c|c|c|}\hline
  $n$&$\alpha(-1)$&$\alpha(0)$&$\alpha(1)$&$\alpha(2)$
  &$\alpha(3)$&$\alpha(4)$&$\alpha(5)$&$\alpha(6)$&$\alpha(7)$\\
  \hline
  1&0&1&0&0&0&0&0&0&0\\
  2&0&3&1&0&0&0&0&0&0\\
  3&0&6&3&1&0&0&0&0&0\\
  4&1&13&8&3&1&0&0&0&0\\
  5&3&24&17&8&3&1&0&0&0\\
  6&10&48&37&19&8&3&1&0&0\\
  7&23&86&71&41&19&8&3&1&0\\
  8&54&161&138&85&43&19&8&3&1\\
  \hline
\end{tabular}
\end{center}

\noindent \underline{$K=2$}
\begin{center}
\begin{tabular}{|c||c|c|c|c|c||c|c|c|c|}\hline
  $n$&$\alpha(-1)$&$\alpha(0)$&$\alpha(1)$&$\alpha(2)$
  &$\alpha(3)$&$\beta(0)$&$\beta(1)$&$\beta(2)$&$\beta(3)$\\
  \hline
  1&0&1&0&0&0&2&0&0&0\\
  2&0&4&1&0&0&10&2&0&0\\
  3&0&12&4&1&0&34&8&2&0\\
  4&1&34&14&4&1&108&30&8&2\\
  \hline
\end{tabular}
\end{center}

\noindent \underline{$K=3$}
\begin{center}
\begin{tabular}{|c||c|c|c|c|c||c|c|c|c|}\hline
  $n$&$\alpha(-1)$&$\alpha(0)$&$\alpha(1)$&$\alpha(2)$
  &$\alpha(3)$&$\beta(0)$&$\beta(1)$&$\beta(2)$&$\beta(3)$\\
  \hline
  1&0&1&0&0&0&2&0&0&0\\
  2&0&5&1&0&0&12&2&0&0\\
  3&0&19&5&1&0&50&10&2&0\\
  \hline
\end{tabular}
\end{center}

\noindent \underline{$K=4$}
\begin{center}
\begin{tabular}{|c||c|c|c|c|c||c|c|c|c|}\hline
  $n$&$\alpha(-1)$&$\alpha(0)$&$\alpha(1)$&$\alpha(2)$
  &$\alpha(3)$&$\beta(0)$&$\beta(1)$&$\beta(2)$&$\beta(3)$\\
  \hline
  1&0&1&0&0&0&2&0&0&0\\
  2&0&6&1&0&0&14&2&0&0\\
  3&0&27&6&1&0&68&12&2&0\\
  \hline
\end{tabular}
\end{center}

\noindent \underline{$K=5$}
\begin{center}
\begin{tabular}{|c||c|c|c|c|c||c|c|c|c|}\hline
  $n$&$\alpha(-1)$&$\alpha(0)$&$\alpha(1)$&$\alpha(2)$
  &$\alpha(3)$&$\beta(0)$&$\beta(1)$&$\beta(2)$&$\beta(3)$\\
  \hline
  1&0&1&0&0&0&2&0&0&0\\
  2&0&7&1&0&0&16&2&0&0\\
  \hline
\end{tabular}
\end{center}

\sectionnew{Appendix B: The Schur function}
\setcounter{section}{2}
\setcounter{subsection}{0}
\setcounter{equation}{0}

The Schur function, which is the character of the general linear
group, can be expressed in terms of free fermions \cite{rS,rDJKM}.
In this appendix we summarize the useful formulae(
\cite{rAFMO3}, see also \cite{rMac}).

\subsection{}
Free fermions\footnote{
We use this notation to avoid a
confusion with the free fermions used
in the free-field realization of $W_{1+\infty}$.
Relation to usual free fermions
 $\bar{\psi}(z)=\sum_{r\in\bz+\frac{1}{2}}
  \bar{\psi}_rz^{-r-\frac{1}{2}}$,
 $\psi(z)=\sum_{r\in\bz+\frac{1}{2}}\psi_rz^{-r-\frac{1}{2}}$ is
 given by $\fb_n=\bar{\psi}_{n+\frac{1}{2}}$,
 $\f_n=\psi_{n-\frac{1}{2}}$.
}
$\fb(z)$, $\f(z)$ and the vacuum state $\dket{0}$ are defined by
\beqa
  &&
  \fb(z)=\sum_{n\in\bz}\fb_nz^{-n-1},\quad
  \f(z)=\sum_{n\in\bz}\f_nz^{-n}, \n
  &&
  \lbrace \fb_m, \f_n \rbrace = \delta_{m+n,0},\quad
  \lbrace \fb_m, \fb_n \rbrace =
  \lbrace \f_m, \f_n \rbrace = 0, \n
  &&
  \fb_m \dket{0} = \f_n \dket{0} =0, \qquad (m\geq 0, n\geq 1).
\eeqa
The fermion Fock space is a linear span of
$\prod_i \fb_{-m_i} \prod_j \f_{-n_j} \dket{0}$.
The $\Uone$ current $\Jc(z)=\sum_{n\in\bz}\Jc_nz^{-n-1}$ is defined by
$\Jc(z)=\;:\fb(z)\f(z):$, {\it i.e.}
$\Jc_n=\sum_{m\in\bz}:\fb_m\f_{n-m}:$,
where the normal ordering $:\fb_m\f_n:$ means
$\fb_m\f_n$ if $m\leq -1$ and $-\f_n\fb_m$ if $m\geq 0$.
Their commutation relations are
\beq
  \lbrack \Jc_n, \Jc_m \rbrack = n\delta_{n+m,0},\quad
  \lbrack \Jc_n, \fb_m \rbrack = \fb_{n+m},\quad
  \lbrack \Jc_n, \f_m \rbrack = -\f_{n+m}.
\eeq
The fermion Fock space is decomposed into the irreducible
representations of $\uone$ with the highest weight state
$\dket{N}$ ($N\in\bZ$),
\beq
  \dket{N}=
  \left\{
  \begin{array}{ll}
    \fb_{-N}\cdots\fb_{-2}\fb_{-1}\dket{0}&N\geq 1 \\
    \dket{0}&N=0 \\
    \f_{N+1}\cdots\f_{-1}\f_{0}\dket{0}&N\leq -1.
  \end{array}
  \right.
  \Label{N}
\eeq

A free boson $\phi(z)$ and the vacuum state $\dket{p}_B$ are
defined by
\beqa
  &&
  \phi(z)
  =
  \hat{q}+\alpha_0\log z-\sum_{n\neq 0}\frac{\alpha_n}{n}z^{-n},\n
  &&
  \lbrack \alpha_n,\alpha_m\rbrack=n\delta_{n+m,0},\quad
  \lbrack \alpha_0,\hat{q}\rbrack=1, \n
  &&
  \alpha_n\dket{p}_B=0\quad(n>0),\quad
  \alpha_0\dket{p}_B=p\dket{p}_B.
\eeqa
The boson Fock space is a linear span of
$\prod_i\alpha_{-n_i}\dket{p}_B$.
The normal ordering $:\quad:$ means that $\alpha_n$ ($n\geq 0$) is
moved to the right of $\alpha_m$ ($m<0$) and $\hat{q}$.
$\dket{p}_B$ is obtained from $\dket{0}$ as $\dket{p}_B
=\;:e^{p\phi(0)}:\dket{p}_B$. The vertex operator satisfies
\beq
  :e^{p\phi(z)}::e^{p'\phi(w)}:\;
  =
  (z-w)^{pp'}:e^{p\phi(z)+p'\phi(w)}:.
\eeq

Boson-fermion correspondence is
\beq
  \fb(z)=\;:e^{\phi(z)}:,\quad
  \f(z)=\;:e^{-\phi(z)}:,\quad
  \dket{N}=\dket{N}_B.
\eeq
$\Uone$ current is $\Jc(z)=\partial\phi(z)$.

A Young diagram has various parametrization:
\beq
\begin{picture}(330,100)
 \put(-20,60){$Y$ $=$}
 \put(10,80){\framebox(80,20){}}
 \put(45,80){\makebox(40,20){$m_1$}}
 \put(10,20){\line(0,1){60}}
 \put(30,20){\line(0,1){60}}
 \put(10,20){\line(1,0){20}}
 \put(10,25){\makebox(20,20){$n_1$}}
 \put(35,70){$\cdot$}
 \put(40,65){$\cdot$}
 \put(45,45){\framebox(40,20){}}
 \put(45,45){\makebox(40,20){$m_h$}}
 \put(45,25){\line(0,1){20}}
 \put(65,25){\line(0,1){20}}
 \put(45,25){\line(1,0){20}}
 \put(45,25){\makebox(20,20){$n_h$}}
 \put(100,60){$=$}
%
 \put(120,80){\framebox(80,20){}}
 \put(120,80){\makebox(40,20){$f_1$}}
 \put(120,40){\line(0,1){40}}
 \put(185,65){$\cdot$}
 \put(175,55){$\cdot$}
 \put(165,45){$\cdot$}
 \put(120,20){\framebox(40,20){}}
 \put(120,20){\makebox(40,20){$f_r$}}
 \put(210,60){$=$}
%
 \put(230,20){\framebox(20,80){}}
 \put(230,60){\makebox(20,40){$g_1$}}
 \put(250,100){\line(1,0){40}}
 \put(260,25){$\cdot$}
 \put(270,35){$\cdot$}
 \put(280,45){$\cdot$}
 \put(290,60){\framebox(20,40){}}
 \put(290,60){\makebox(20,40){$g_c$}}
\end{picture}
  \Label{Young}
\eeq
where $m_1>\cdots>m_h\geq 1$, $n_1>\cdots>n_h\geq 0$,
$f_1\geq\cdots\geq f_r\geq 1$,
$g_1\geq\cdots\geq g_c\geq 1$.
According to these parametrizations, we denote the Young diagram $Y$ by
$Y=(m_1,\cdots,m_h;n_1,\cdots,n_h)$,
$Y=(f_1,\cdots,f_r)$ or ${}^tY=(g_1,\cdots,g_c)$ respectively,
and the number of boxes as
$|Y|=\sum_{i=1}^h(m_i+n_i)=\sum_{i=1}^rf_i=\sum_{i=1}^cg_i$.
Corresponding to the Young diagram \eq{Young}, we define a state
$\dket{N;Y}$ as follows:
\beqa
  \dket{N;Y}
  &\!\!=\!\!&
  \prod_{i=1}^h\fb_{-m_i-N}\f_{-n_i+N} (-1)^{n_i} \dket{N}
  \Label{NYh} \\
  &\!\!=\!\!&
  \fb_{-\bar{f}_1-N}\fb_{-\bar{f}_2-N}\cdots\fb_{-\bar{f}_r-N}
  \dket{N-r}
  \Label{NYr} \\
  &\!\!=\!\!&
  (-1)^{|Y|}\f_{-\bar{g}_1+N}\f_{-\bar{g}_2+N}\cdots\f_{-\bar{g}_r+N}
  \dket{N+c},
  \Label{NYc}
\eeqa
where
\beq
  \bar{f}_i=f_i-i+1,\quad
  \bar{g}_i=g_i-i.
\eeq
These states $\dket{N;Y}$ with all Young diagrams are a basis of
$\uone$ representation space of the highest weight $\dket{N}$.
We abbreviate $\dket{0;Y}$ as $\dket{Y}$.

Bra states are obtained from ket states by $\dagger$ operation
($\fb_n{}^{\dagger}=\f_{-n}$) with the normalization
$\langle\!\langle0\dket{0}=1$; for example,
$\dbra{N}=\dket{N}^{\dagger}$ and
$\langle\!\langle N\dket{N'}=\delta_{NN'}$,
$\dbra{Y}=\dket{Y}^{\dagger}=\dbra{0}
\prod_{i=1}^h \fb_{n_i}\f_{m_i} (-1)^{n_i}$ and
$\langle\!\langle Y\dket{Y'}=\delta_{YY'}$.
Note that $\{ \dket{N;Y} \}$ is an orthonormal basis of the fermion
Fock space with $\Uone$--charge $N$.

Irreducible representations of the permutation group $\Sn$ and the
general linear group $GL(N)$ are both characterized by the Young
diagrams $Y$.
We denote their characters by $\chi_Y(k)$ and $\tau_Y(x)$, respectively.
Here $(k)=1^{k_1}2^{k_2}\cdots n^{k_n}$ stands for the conjugacy class
of $\Sn$; $k_1+2k_2+\cdots+nk_n=n=$the number of boxes in $Y$.
$x=[x_{\ell}]$ ($\ell=1,2,3,\cdots$) stands for
$x_{\ell}=\frac{1}{\ell}\mbox{tr }g^{\ell}
=\frac{1}{\ell}\sum_{i=1}^N\epsilon_i^{\ell}$ for an element $g$ of
$GL(N)$ whose diagonalized form is
$g={\rm diag}[\epsilon_1,\epsilon_2,\cdots,\epsilon_N]$.
In this case the number of boxes in $Y$ is a rank of
tensor for $GL(N)$.
We take $N\rightarrow\infty$ limit formally.
$\tau_Y$ is called the Schur function.
The skew S-function $\tau_{Y/Y'}$ is defined by
\beq
  \tau_{Y/Y'}(x)
  =
  \sum_{Y''}C^Y_{Y'Y''}\tau_{Y''}(x),
\eeq
where the Clebsch-Gordan coefficients $C^{Y}_{Y'Y''}$ are
\beq
  \tau_{Y'}(x)\tau_{Y''}(x)
  =
  \sum_YC^Y_{Y'Y''}\tau_Y(x),
\eeq
namely decomposition of the tensor product of representations $Y'$
and $Y''$; $Y'\otimes Y''=\bigoplus_{Y}C^{Y}_{Y'Y''}Y$.
$\tau_{Y/Y'}(x)$ is non-vanishing only for $Y'\subseteq Y$.

$\chi_Y(k)$, $\tau_Y(x)$ and $\tau_{Y/Y'}(x)$ are expressed in
terms of free fermion as follows:
\beqa
  \chi_Y(k) &\!\!=\!\!&
  \dbra{0} \Jc_1^{k_1}\Jc_2^{k_2}\cdots\Jc_n^{k_n} \dket{Y},
  \Label{chiY} \\
  \tau_Y(x) &\!\!=\!\!&
  \dbra{0}\exp\biggl(\sum_{\ell=1}^{\infty}x_{\ell}\Jc_{\ell}\biggr)
  \dket{Y},
  \Label{tauY} \\
  \tau_{Y/Y'}(x) &\!\!=\!\!&
  \dbra{Y'}\exp\biggl(\sum_{\ell=1}^{\infty}x_{\ell}\Jc_{\ell}\biggr)
  \dket{Y}.
  \Label{tauYY'}
\eeqa
We remark that they can also be written as
$\chi_Y(k)= \dbra{Y}
\Jc_{-1}^{k_1}\Jc_{-2}^{k_2}\cdots\Jc_{-n}^{k_n} \dket{0}$,
$\tau_Y(x)=\dbra{Y} \exp(
\sum_{\ell=1}^{\infty} x_{\ell}\Jc_{-\ell})\dket{0}$ and
$\tau_{Y/Y'}(x)=\dbra{Y} \exp(
\sum_{\ell=1}^{\infty} x_{\ell}\Jc_{-\ell})\dket{Y'}$.

Under the adjoint action of
$\exp(\sum_{\ell=1}^{\infty} x_{\ell}\Jc_{\ell})$,
$\fb(z)$ and $\f(z)$ transform as
\beqa
  \exp\biggl(\sum_{\ell=1}^{\infty}x_{\ell}\Jc_{\ell}\biggr)
  \fb(z)
  \exp\biggl(-\sum_{\ell=1}^{\infty}x_{\ell}\Jc_{\ell}\biggr)
  &\!\!=\!\!&
  \exp\biggl(\sum_{\ell=1}^{\infty}x_{\ell}z^{\ell}\biggr)
  \fb(z), \n
  \exp\biggl(\sum_{\ell=1}^{\infty}x_{\ell}\Jc_{\ell}\biggr)
  \f(z)
  \exp\biggl(-\sum_{\ell=1}^{\infty}x_{\ell}\Jc_{\ell}\biggr)
  &\!\!=\!\!&
  \exp\biggl(-\sum_{\ell=1}^{\infty}x_{\ell}z^{\ell}\biggr)
  \f(z).
\eeqa

\subsection{}

Let us introduce the elementary Schur polynomial $P_n(x)$,
\beq
  \exp\biggl(\sum_{\ell=1}^{\infty}x_{\ell}z^{\ell}\biggr)
  =
  \sum_{n\in\bz}P_n(x)z^n.
  \Label{Pn}
\eeq
Note that $P_n(x)=0$ for $n<0$.
Then the Schur functions with one row, one column and one hook
are respectively given by
\beqa
  \tau_f(x)&\!\!=\!\!&P_f(x), \\
  \tau_{1^g}(x)&\!\!=\!\!&(-1)^gP_g(-x), \\
  \tau_{m;n}(x)&\!\!=\!\!&
  (-1)^n\sum_{\ell=0}^{\infty}P_{m+\ell}(x)P_{n-\ell}(-x) \n
  &\!\!=\!\!&
  (-1)^{n-1}\sum_{\ell=0}^{\infty}P_{m-1-\ell}(x)P_{n+1+\ell}(-x).
  \Label{taumn}
\eeqa
Using this, the Schur function with the Young diagram \eq{Young}
is given by
\beqa
  \tau_{Y}(x)
  &\!\!=\!\!&
  \det\Bigl(\tau_{m_i;n_j}(x)\Bigr)_{1\leq i,j\leq h}
  \Label{tauh} \\
  &\!\!=\!\!&
  \det\Bigl(P_{f_i-i+j}(x)\Bigr)_{1\leq i,j\leq r}
  \Label{taur} \\
  &\!\!=\!\!&
  (-1)^{|Y|}\det\Bigl(P_{g_i-i+j}(-x)\Bigr)_{1\leq i,j\leq c}.
  \Label{tauc}
\eeqa
The Schur function with the transposed Young diagram ${}^tY$ is
\beq
  \tau_{{}^tY}(x)=(-1)^{|Y|}\tau_Y(-x).
\eeq
The skew S-function with Young diagrams parametrized in the second and
third form of \eq{Young}, is given by
\beqa
  \tau_{Y/Y'}(x)
  &\!\!=\!\!&
  \det\Bigl(P_{f_i-f'_j-i+j}(x)\Bigr)_{1\leq i,j\leq r} \\
  &\!\!=\!\!&
  (-1)^{|Y|-|Y'|}\det\Bigl(P_{g_i-g'_j-i+j}(-x)\Bigr)_{1\leq i,j\leq c}.
\eeqa

Since $\tau_Y$'s are a basis of the space of symmetric functions,
we have
\beq
  \prod_r\prod_s \frac{1}{1-u_rv_s}
  =
  \sum_Y \tau_Y(x)\tau_Y(y),
  \label{YY}
\eeq
where the summation runs over all the Young diagrams and $x,y$ are
the Miwa variables for $u,v$,
\beq
  x_{\ell}=\frac{1}{\ell}\sum_ru_r^{\ell},\quad
  y_{\ell}=\frac{1}{\ell}\sum_sv_s^{\ell}.
\eeq
Similarly we have
\beq
  \sum_Y\tau_{Y/Y'}(x)\tau_Y(y)
  =
  \sum_Y\tau_Y(x)\tau_Y(y)\tau_{Y'}(y).
\eeq

Those formulae are easily proved by using free field expression
\eqs{tauY}{tauYY'} and bosonization.
For example, \eq{taur} is obtained as follows.
By rewriting $\dket{Y}$ \eq{NYr} as
$$
  \dket{Y}
  =
  \oint\prod_{i=1}^r\frac{dz_i}{2\pi i}z_i^{-\bar{f}_i}
  \cdot
  \fb(z_1)\cdots\fb(z_r)\dket{-r},
$$
\eq{tauY} becomes
$$
  \tau_Y(x)
  =
  \oint\prod_{i=1}^r\frac{dz_i}{2\pi i}z_i^{-\bar{f}_i}
  e^{\sum_{\ell=1}^{\infty}x_{\ell}z_i^{\ell}}
  \cdot
  \dbra{0}\fb(z_1)\cdots\fb(z_r)\dket{-r}.
$$
Bosonization tells us that
$$
  \dbra{0}\fb(z_1)\cdots\fb(z_r)\dket{-r}
  =
  \prod_{i<j}(z_i-z_j) \cdot \prod_{i=1}^rz_i^{-r}.
$$
Since $\prod_{i<j}(z_i-z_j)$ is the Vandermonde determinant
$(-1)^{\frac{1}{2}r(r-1)}\det\bigl(z_i^{j-1}\bigr)_{1\leq i,j\leq r}$,
we obtain \eq{taur} after picking up residues.\\
Eq.\,(\ref{YY}) is proved as follows:
\beqan
  &&
  \prod_r\prod_s \frac{1}{1-u_rv_s}
  =
  \exp\biggl(\sum_r\sum_s\log\frac{1}{1-u_rv_s}\biggr)=
  \exp\biggl(\sum_r\sum_s
  \sum_{\ell=1}^{\infty}\frac{1}{\ell}(u_rv_s)^{\ell}\biggr) \\
  &&
  =
  \exp\biggl(\sum_{\ell=1}^{\infty}\ell x_{\ell}y_{\ell}\biggr)
  =
  \dbra{0}\exp\biggl(\sum_{\ell=1}^{\infty}
  x_{\ell}\Jc_{\ell}\biggr)
  \exp\biggl(\sum_{\ell=1}^{\infty}
  y_{\ell}\Jc_{-\ell}\biggr)\dket{0} \\
  &&
  =
  \sum_Y
  \dbra{0}\exp\biggl(\sum_{\ell=1}^{\infty}
  x_{\ell}\Jc_{\ell}\biggr)
  \dket{Y}\dbra{Y}
  \exp\biggl(\sum_{\ell=1}^{\infty}y_{\ell}
  \Jc_{-\ell}\biggr)\dket{0}
  =
  \sum_Y\tau_Y(x)\tau_Y(y).
\eeqan
Here we have used the completeness of $\{ \dket{Y} \}$ in the
fermion Fock space with vanishing $\Uone$--charge.

\subsection{}

In this subsection we set $x_{\ell}$ as follows:
\beq
  x_{\ell}=\frac{1}{\ell}\frac{q^{a\ell}}{1-q^{\ell}}.
\eeq
Then the Schur functions with one row and one column are
given by
\beqa
  P_n(x)
  &\!\!=\!\!&
  \prod_{j=1}^n\frac{q^a}{1-q^j},
  \Label{qPnr} \\
  (-1)^nP_n(-x)
  &\!\!=\!\!&
  \prod_{j=1}^n\frac{q^{j-1+a}}{1-q^j}.
  \Label{qPnc}
\eeqa
{}From \eq{taumn}, the Schur function with one hook becomes
\beq
  \tau_{m;n}(x)
  =
  q^{a(m+n)+\frac{1}{2}n(n+1)}
  \prod_{j=1}^{m-1}\frac{1}{1-q^j}
  \prod_{j=1}^{n}\frac{1}{1-q^j}
  \cdot\frac{1}{1-q^{m+n}}.
\eeq
By combining those and eqs.\,(\ref{tauh},\ref{taur},\ref{tauc}),
the Schur function with the Young diagram \eq{Young} is given by
\beqa
  \tau_Y(x)
  &\!\!=\!\!&
  q^{a|Y|
     +\sum_{i=1}^h\bigl(\frac{1}{2}n_i(n_i+1)+(i-1)(m_i+n_i)\bigr)} \n
  &&\times
  \prod_{i=1}^h\biggl(
    \prod_{j=1}^{m_i-1}\frac{1}{1-q^j}
    \prod_{j=1}^{n_i}\frac{1}{1-q^j}\biggr)
  \cdot
  \frac{\prod_{i<j}(1-q^{m_i-m_j})(1-q^{n_i-n_j})}
       {\prod_{i,j}(1-q^{m_i+n_j})} \\
  &\!\!=\!\!&
  q^{a|Y|+\sum_{i=1}^r(i-1)f_i}
  \prod_{i=1}^r\prod_{j=1}^{f_i-i+r}\frac{1}{1-q^j}
  \cdot
  \prod_{i<j}(1-q^{f_i-f_j-i+j})
  \Label{qtaur} \\
  &\!\!=\!\!&
  q^{a|Y|+\sum_{i=1}^r\frac{1}{2}g_i(g_i-1)}
  \prod_{i=1}^c\prod_{j=1}^{g_i-i+c}\frac{1}{1-q^j}
  \cdot
  \prod_{i<j}(1-q^{g_i-g_j-i+j}).
  \Label{qtauc}
\eeqa


\end{document}